\documentclass[10pt,twoside,letter,amsmath,amssymb,aps,showkeys,showpacs]{revtex4}
\usepackage{amsfonts}
\usepackage{graphics}
\usepackage{epsfig}
\usepackage{color}

\begin{document}

\thispagestyle{myheadings}

\title{
Electroweak radiative corrections for polarized M{\o}ller scattering \\ at 
future 11 GeV JLab experiment}

\author{Aleksandrs Aleksejevs}
\email{aaleksejevs@swgc.mun.ca}
\affiliation{Memorial University, Corner Brook, Canada}

\author{Svetlana Barkanova}
\email{svetlana.barkanova@acadiau.ca}
\affiliation{Acadia University, Wolfville, Canada}

\author{Alexander Ilyichev}
\email{ily@hep.by}
\affiliation{
National Center of Particle and High Energy Physics, Minsk, Belarus}

\author{Vladimir Zykunov}
\email{vladimir.zykunov@cern.ch}
\affiliation{Belarussian State University of Transport, Gomel, Belarus}

\begin{abstract}
We perform updated and detailed calculations of the complete NLO set of electroweak 
radiative corrections to parity violating $e^-e^- \rightarrow e^-e^- (\gamma)$ 
scattering asymmetries at energies relevant for the ultra-precise M{\o}ller experiment 
coming soon at JLab. 
Our numerical results are presented for a range of experimental cuts and relative importance 
of various contributions is analyzed. We also provide very compact expressions analytically 
free from non-physical parameters and show them to be valid for fast yet accurate estimations.
\end{abstract}


\maketitle

\section{Introduction}

There are several reasons why polarized M{\o}ller scattering has attracted so much interest 
from both experimental and theoretical communities. 
The scattering of two identical polarized fermions has been used in many laboratories 
for the high-precision determination of the electron-beam polarization, including SLC 
in \cite{11}, E-143 \cite{12} and E-154 \cite{13} experiments at SLAC, and several 
experiments at JLab \cite{14} and MIT-Bates \cite{15}. A M{\o}ller polarimeter may 
also be of use in future experiments planned at ILC \cite{18}.
An experiment E-158 at SLAC \cite{2}, which studied M{\o}ller scattering of 45- to 48-GeV 
polarized electrons on unpolarized electrons of a hydrogen target, allowed one of the important 
parameters in the Standard Model - sine of the  Weinberg angle - to be determined with an 
unprecedented accuracy. The new-generation experiment at 11 GeV soon to be started at JLab 
will offer a new level of sensitivity and measure the parity-violating asymmetry in the scattering 
of longitudinally polarized electrons off unpolarized electrons to a precision of 0.73 ppb. 
That would allow a measurement of the weak charge of the electron to a fractional accuracy 
of 2.3\% and a determination of the weak mixing angle with an uncertainty of 
$\pm$ 0.00026 (stat.) $\pm$ 0.00013 (syst.) \cite{JLab12}.

Since $e^-e^- \rightarrow e^-e^- (\gamma)$ scattering is a very clean process with well-known 
initial energy and  extremely suppressed backgrounds, any inconsistency with the Standard Model 
will signal new physics. M{\o}ller scattering experiments can provide indirect access to physics 
at multi-TeV scales and play important complementary role to the LHC research program \cite{1}. 

Obviously, before we can extract reliable information from experimental data, it is necessary 
to take into account higher order effects. These are the processes which are more complicated 
than the process being studied, but which are indistinguishable from it experimentally. 
This procedure of the inclusion of radiative corrections is an indispensable part of any modern 
experiment, but will be of the most paramount importance for the ultra-precise measurement of the 
weak mixing angle via 11 GeV M{\o}ller scattering planned at JLab.

One of the earliest works on Electroweak Radiative Corrections (EWC) to the observables 
of M{\o}ller scattering was done by Czarnecki and Marciano \cite{4-CzMa}. According to 
their calculations for the asymmetry in the E-158 kinematical region, the EWC reduce the 
tree level prediction by 40$\pm$3\%. 
A close value of the asymmetry was obtained in the 
study of Denner and Pozzorini \cite{5-DePo}, where radiative corrections in polarized 
M{\o}ller scattering were examined at arbitrary, including high, energies. It is worth 
noticing that these two studies used different renormalization schemes: \cite{4-CzMa} 
employed the modified minimal-subtraction scheme and \cite{5-DePo} used the on-shell 
renormalization. However, the authors of \cite{4-CzMa} and \cite{5-DePo} did not take 
into account all radiative contributions consistently. For example, they completely 
disregarded the Hard Photon Bremsstrahlung (HPB) contribution assuming that it was small.
The first computation of the HPB contributions with realistic experimental cuts was done 
by Petriello \cite{6-Pe} who studied a total set of the lowest-order EWC under kinematical 
conditions of E-158 experiment. He clearly showed that experimental cuts play a significant role.
Later works \cite{7}, \cite{8} and \cite{9} dedicated to the EWC for the E-158 experiment 
which employed a covariant method for extracting of infrared divergences \cite{10} showed 
a good agreement with \cite{4-CzMa} for the bulk of weak-interaction contributions and 
proved that the effect of HPB was below the experimental error of E-158.

	Electromagnetic radiative corrections, which dominate weak interaction
effects at the energies of E-158, and their effect on the measurement of beam 
polarization in a M{\o}ller polarimeter were assessed in \cite{19}. A different 
calculation of these electromagnetic corrections was undertaken in \cite{20}, 
where attention was primarily given to the effect of experimental cuts on the 
inelasticity and, accordingly, proving the need for taking into account 
hard-photon bremsstrahlung in experiments of this kind. A Monte Carlo generator 
that makes it possible to simulate radiative events in a M{\o}ller process at 
moderate energies with the aim of determining beam polarization was developed in \cite{21}.

	Finally, an attempt to calculate the hard bremsstrahlung contribution for M{\o}ller 
process at high energies was made in paper  \cite{36}.
A feature specific to this calculation, as well as to the calculations in \cite{5-DePo} and 
\cite{6-Pe}, was the use of a  parameter $\omega$ that separates regions of soft and hard photons. 
As a result, the Monte Carlo integration technique permitted the development of a code which made 
it possible to take into account radiative corrections  in the case of arbitrary cuts imposed on 
the detection of electrons. 

	Although obviously significant theoretical effort has been already dedicated to this task, 
we can see at least three major reasons for addressing EWC to M{\o}ller scattering yet again. 
First, since the new experiment at Jlab will provide more precise  information 
than, for example, E-158, the theoretical predictions for this ultra-precise measurement 
must be made with a full treatment of one-loop radiative corrections and at least leading 
two-loop corrections. In this work, we calculated the full set of the  one-loop corrections 
both numerically with no simplifications using FeynArts and FormCalc \cite{Hahn} as the base languages 
and analytically in a compact form useful for fast estimations. This way, we can control 
an accuracy of asymptotic approximations at the one-loop level very well now, so we hope 
that the same idea will help us to control our accuracy as we calculate leading corrections 
at two-loop level in the nearest future. 
The next major goal of the presented paper is to get EWC in a form which is analytically free 
from nonphysical parameters. 
And, finally, we believe that the complexity of the problem demands the tuned comparison with 
different calculation schemes. This will be our next task.

The rest of the paper is organized as follows. The basic notations as well as
the lowest-order  (Born) contribution to Moller scattering are presented in Section II. 
The contribution from a full set of additional virtual particles (such as self-energies, vertices and boxes) 
is described in Section III. 
The real photon emission and its separation on hard and soft parts as well as
cancellation of the  infrared divergences  together 
with the  unphysical parameters  can be found in Section IV. 
The numerical analysis is presented in Section V and Appendix B.

\section{BORN CROSS SECTION: BASIC NOTATION AND CONVENTIONS}

\begin{figure}
\vspace{10mm}
\begin{tabular}{cc}
\begin{picture}(60,60)
\put(-150,-60){
\epsfxsize=7cm
\epsfysize=7cm
\epsfbox{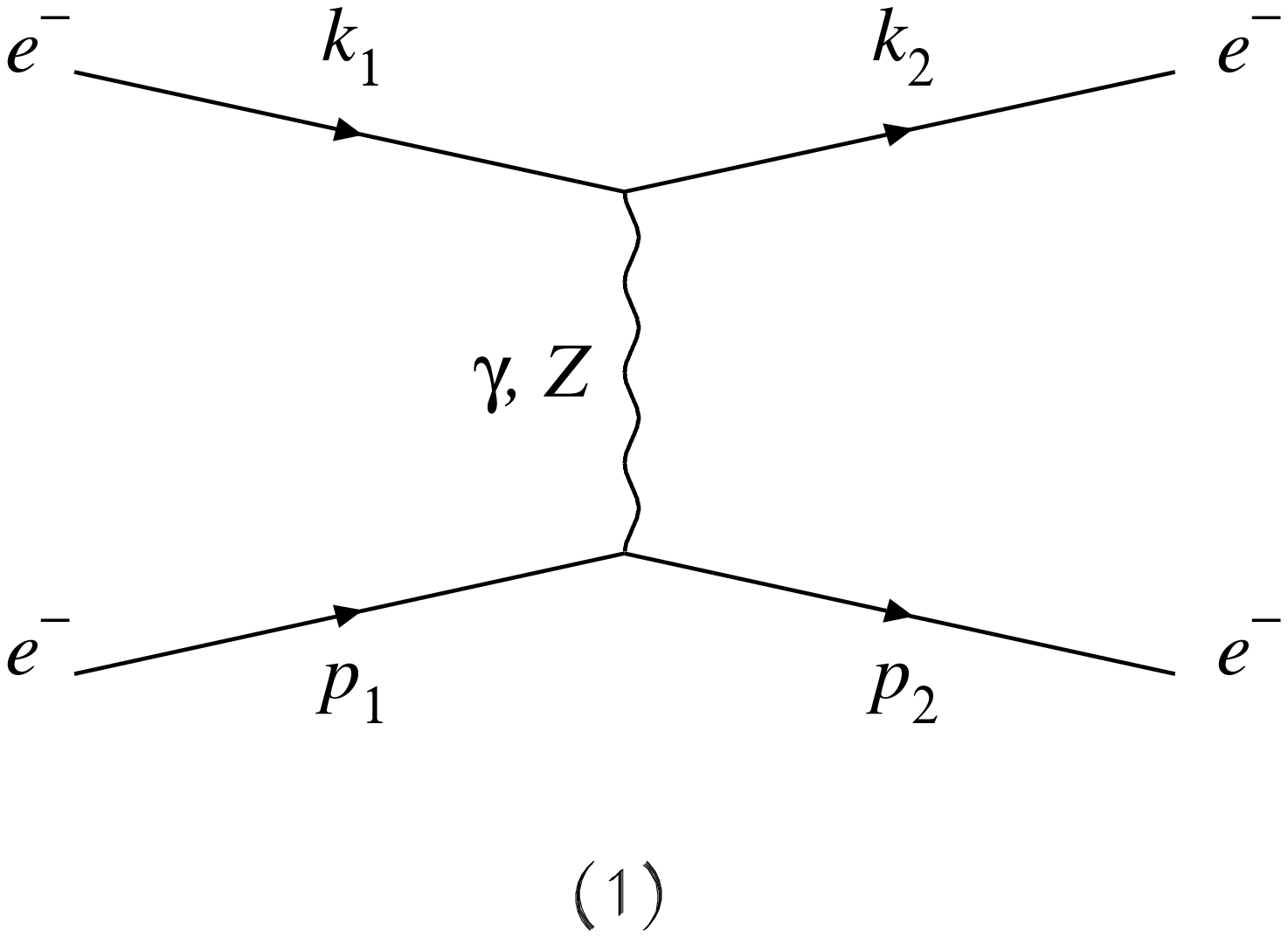} }
\end{picture}
&
\begin{picture}(60,100)
\put(-20,-60){
\epsfxsize=7cm
\epsfysize=7cm
\epsfbox{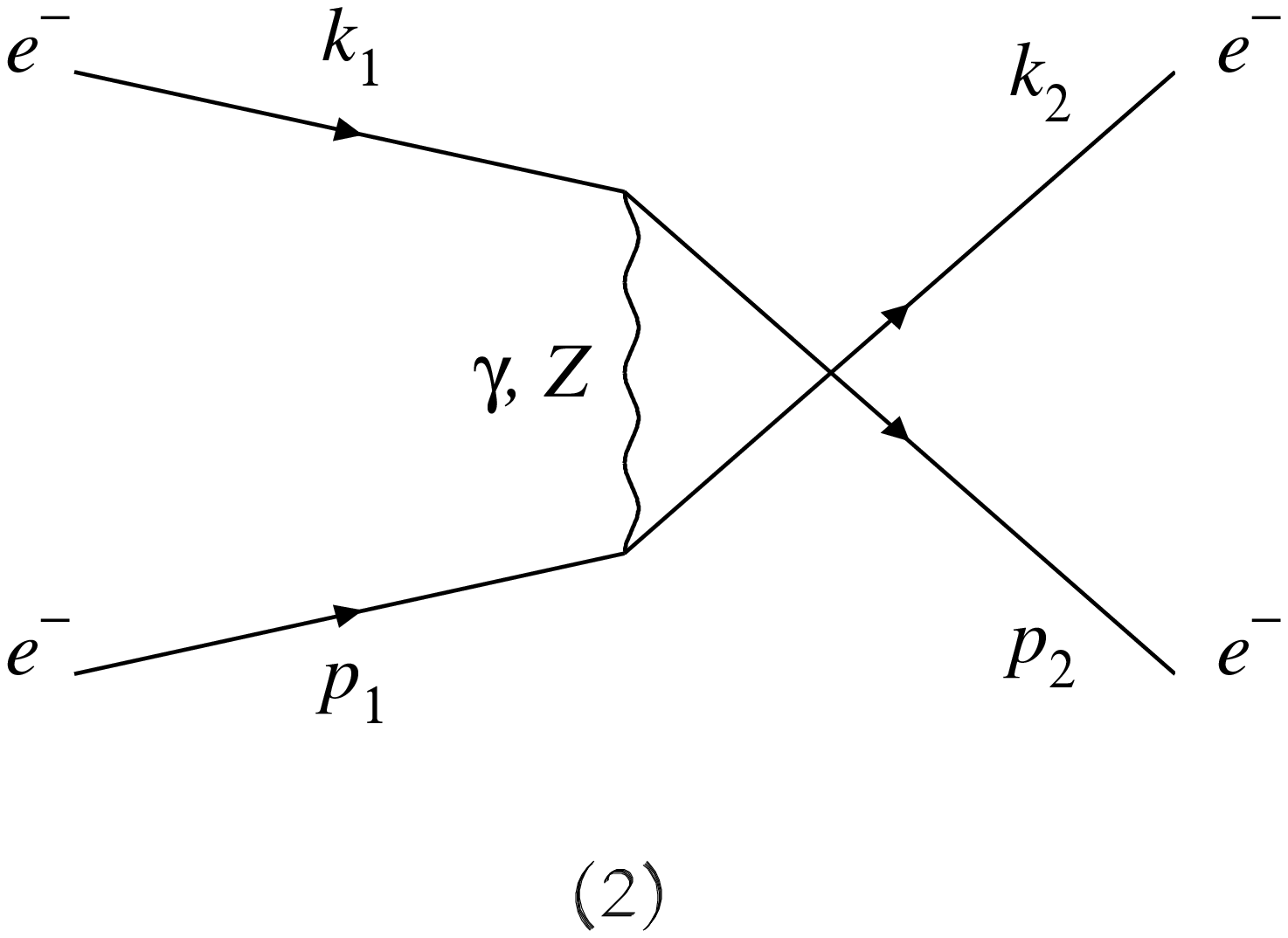} }
\end{picture}
\end{tabular}
\vspace{10mm}
\caption{\protect\it 
Diagrams describing nonradiative M{\o}ller scattering in the (1) t- and (2) u-channels.}
\label{born}
\vspace{5mm}
\end{figure}

In the Standard Model, the Born cross section for
the M{\o}ller scattering 
\begin{equation}
e^-(k_1,\xi)+e^-(p_1,\eta) \rightarrow e^-(k_2)+e^-(p_2)
\label{0} 
\end{equation}
can be represented in the form
\begin{equation}
\sigma^0 =\frac{\pi \alpha^2}{s}
\sum_{i,j=\gamma,Z} [\lambda_-^{i,j}(u^2D^{it}D^{jt}+t^2D^{iu}D^{ju})
   + \lambda_+^{i,j}s^2(D^{it}+D^{iu})(D^{jt}+D^{ju})],
\label{cs0} 
\end{equation}
with
$\sigma \equiv {d\sigma}/{d \cos \theta}$ where
$\theta$  is the scattering angle of the detected electron
with the 4-momentum $k_2$ in the center of mass system of the initial electrons and
the 4-momenta of initial ($k_1$ and $p_1$) and final
($k_2$ and $p_2$) electrons (see Fig.~\ref{born}) generate a standard
set of Mandelstam variables,
\begin{equation}
s=(k_1+p_1)^2,\ t=(k_1-k_2)^2,\ u=(k_2-p_1)^2,
\label{stu}
\end{equation}
while the electron polarization vectors $\xi$ and $\eta$ have
the form \cite{ash}:
\begin{equation}
\xi \approx \frac{k_1}{m}-\frac{2m}{s}p_1, \ \
\eta \approx \frac{2m}{s}k_1-\frac{p_1}{m}.
\label{vec}
\end{equation}
It should also be noted that the electron mass $m$ is
disregarded wherever possible, in particular if
$m^2 \ll s,-t,-u$.

A useful structure in the present study is
\begin{equation}
D^{il}=\frac{1}{l-m_i^2}\ \ (i=\gamma,Z;\ l=t,u),
\label{structure}
\end{equation}
which depends on the $Z$-boson mass $m_Z$ and
on the photon mass $m_\gamma$. The photon mass  is set to zero
everywhere with the exception of specially indicated
cases where the photon mass is taken to be an
infinitesimal parameter that regularizes an infrared
divergence. Another set of useful functions is
\begin{eqnarray}
{\lambda_{\pm}}^{i,k} =
     {\lambda_1}_B^{i,k}{\lambda_1}_T^{i,k} \pm {\lambda_2}_B^{i,k}{\lambda_2}_T^{i,k},
\label{b10}
\end{eqnarray}
which are combinations of coupling constants
and $p_{B(T)}$, where $p_{B(T)}$ are the degrees of polarizations
of electrons with 4-momentum $k_1$ ($p_1$).
Specifically, they are given by
\begin{eqnarray}
 {\lambda_1}_{B(T)}^{i,j} = \lambda_V^{i,j} -p_{B(T)} \lambda_A^{i,j},\
   {\lambda_2}_{B(T)}^{i,j} = \lambda_A^{i,j} -p_{B(T)} \lambda_V^{i,j},
\nonumber
\end{eqnarray}
\begin{equation}
 \lambda_V^{i,j}=v^iv^j + a^ia^j,\
   \lambda_A^{i,j}=v^ia^j + a^iv^j,
\end{equation}
where
\begin{equation}
 v^{\gamma}=1,\ a^{\gamma}=0,\
 v^Z=(I_e^3+2s_{W}^2)/(2s_{W}
c_{W}),
\ a^Z=I_e^3/(2s_{W}c_{W}).
\end{equation}
It should be recalled that $ I_e^3=-1/2 $ and 
$s_{W}\  (c_{W})$ 
are the sine (cosine) of the Weinberg angle expressed in terms of the $Z$- and $W$-boson
masses according to the rules of the Standard Model:
\begin{equation}
c_{W}=m_{W}/m_{Z},\
s_{W}=\sqrt{1-c_{W}^2}.
\end{equation}

The electron polarization degrees $p_{B(T)}$ in the
cross sections are labeled here as follows: 
the subscripts
$L$ and $R$ on the cross sections correspond
to $p_{B(T)}$ = $-1$ and $p_{B(T)}$ = $+1$, where the first subscript
indicates the degree of polarization for the 
4-momentum $k_1$, while the second one indicates the
degree of polarization for the 4-momentum $p_1$. By
combining the degrees of electron beam polarizations,
we can obtain four measurable cross sections,
but, by virtue of the rotational invariance, the two of them will be
identical: $\sigma_{LR}=\sigma_{RL}$. From the three cross sections
we can construct three independent asymmetries \cite{CC96}
(which are very close at large scattering
angles), and the two of them (mainly $A_1$) are main subject of our investigation:
\begin{equation}
A_1 =
 \frac{\sigma_{LL}+\sigma_{LR}-\sigma_{RL}-\sigma_{RR}}
      {\sigma_{LL}+\sigma_{LR}+\sigma_{RL}+\sigma_{RR}}
 =
 \frac{\sigma_{LL}-\sigma_{RR}}
      {\sigma_{LL}+2\sigma_{LR}+\sigma_{RR}},
\label{A}
\end{equation}
\begin{equation}
A_2 =
 \frac{\sigma_{LL}-\sigma_{RR}}
      {\sigma_{LL}+\sigma_{RR}}.
\label{A2}
\end{equation}

All of the asymmetries are proportional to the
combination $1-4s_{W}^2$ (by virtue of the proportionality
of the cross-section difference $\sigma_{LL}-\sigma_{RR}$) and are
therefore highly sensitive to small changes in $s_{W}$.
This is precisely the reason why the asymmetry $A_1$,
which, at moderately low energies, is given by
\begin{equation}
A_1 = \frac{s}{2m_{W}^2}
\frac{y(1-y)}{1+y^4+(1-y)^4} \frac{1-4s_{W}^2}{s_{W}^2},\ \ y=-t/s,
\end{equation}
was used as the observable in the E-158
and will be measured in the future experiment at JLab with 11 GeV electrons.

In addition to a fact that the asymmetry $A_1$ is a
parity-violating observable, 
this asymmetry has yet another remarkable property
based on its structure: it only involves combinations
$\sigma_{LL}+\sigma_{LR}$ and $\sigma_{RL}+\sigma_{RR}$, 
which can be
interpreted as the cross sections for the scattering
of electrons having the polarizations of $p_B=-1$ and 
$p_B=+1$ on unpolarized electrons. 
Frequently, $A_1$ is said to be a single-polarization asymmetry, in contrast
to the asymmetry constructed as 
\begin{equation}
A_{LR} = \frac{\sigma_{LR}-\sigma_{LL}}{\sigma_{LR}+\sigma_{LL}},
\end{equation}
which consists of the cross sections for
the scattering of electrons having different polarizations.
The latter asymmetry, which conserves parity,
is the most important observable in determining the
electron-beam polarization with the aid of a M{\o}ller
polarimeter.

\section{EWC: Contribution of Additional Virtual Particles}

\subsection{ Boson Self-Energies}
\begin{figure}
\vspace{15mm}
\begin{tabular}{ccc}
\begin{picture}(60,60)
\put(-120,0){
\epsfxsize=5cm
\epsfysize=5cm
\epsfbox{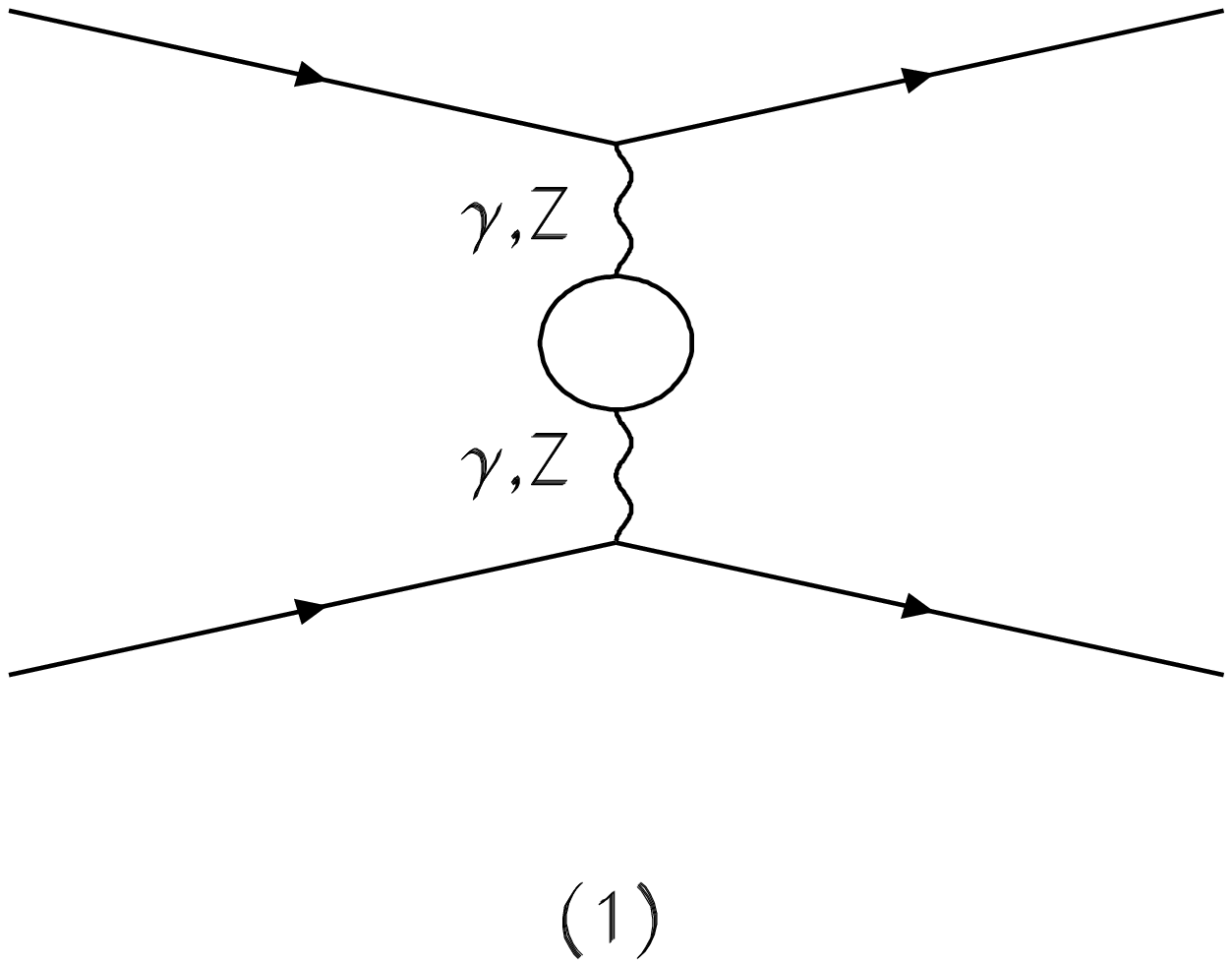} }
\end{picture}
&
\begin{picture}(60,100)
\put(-70,0){
\epsfxsize=5cm
\epsfysize=5cm
\epsfbox{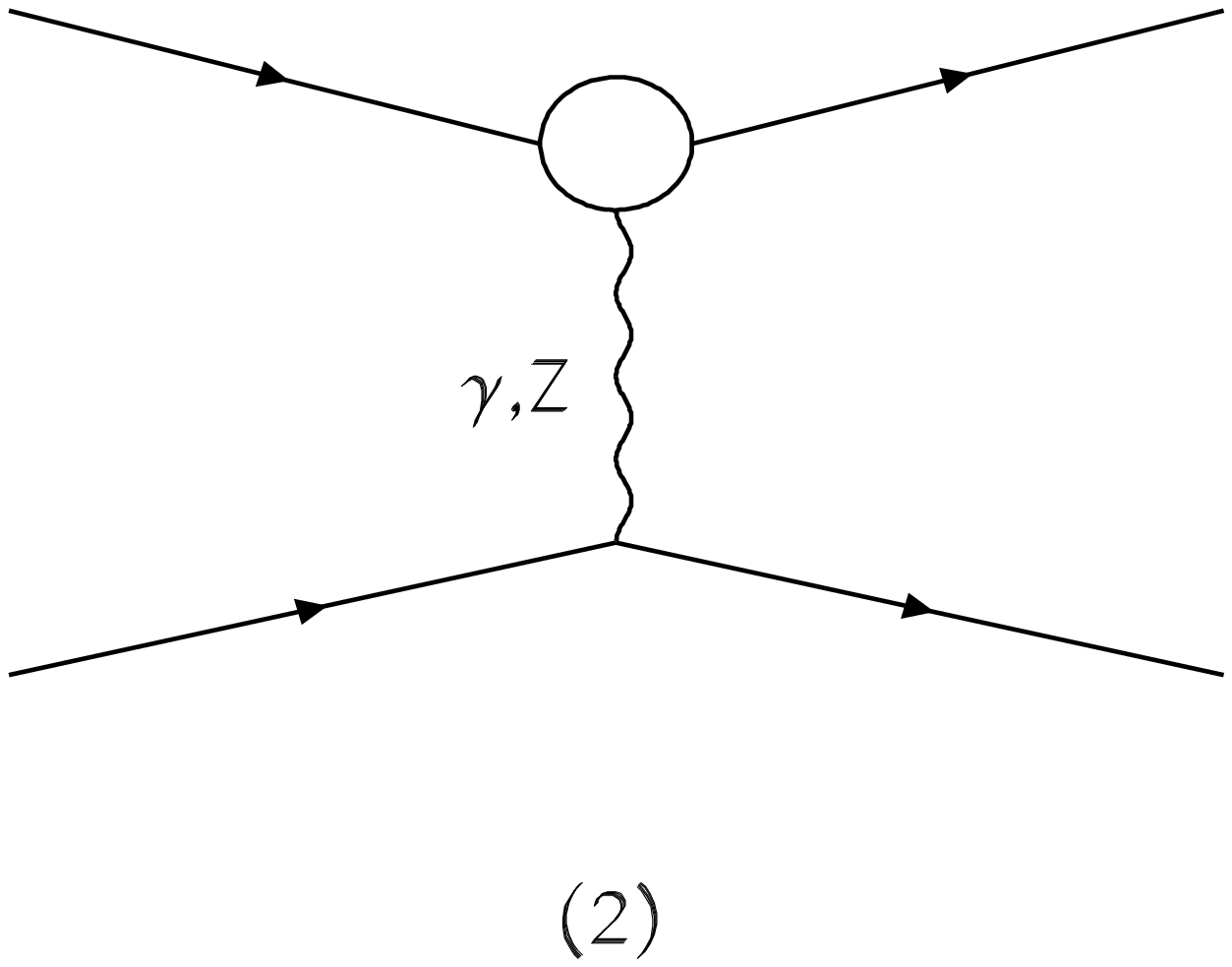} }
\end{picture}
&
\begin{picture}(60,100)
\put(-20,0){
\epsfxsize=5cm
\epsfysize=5cm
\epsfbox{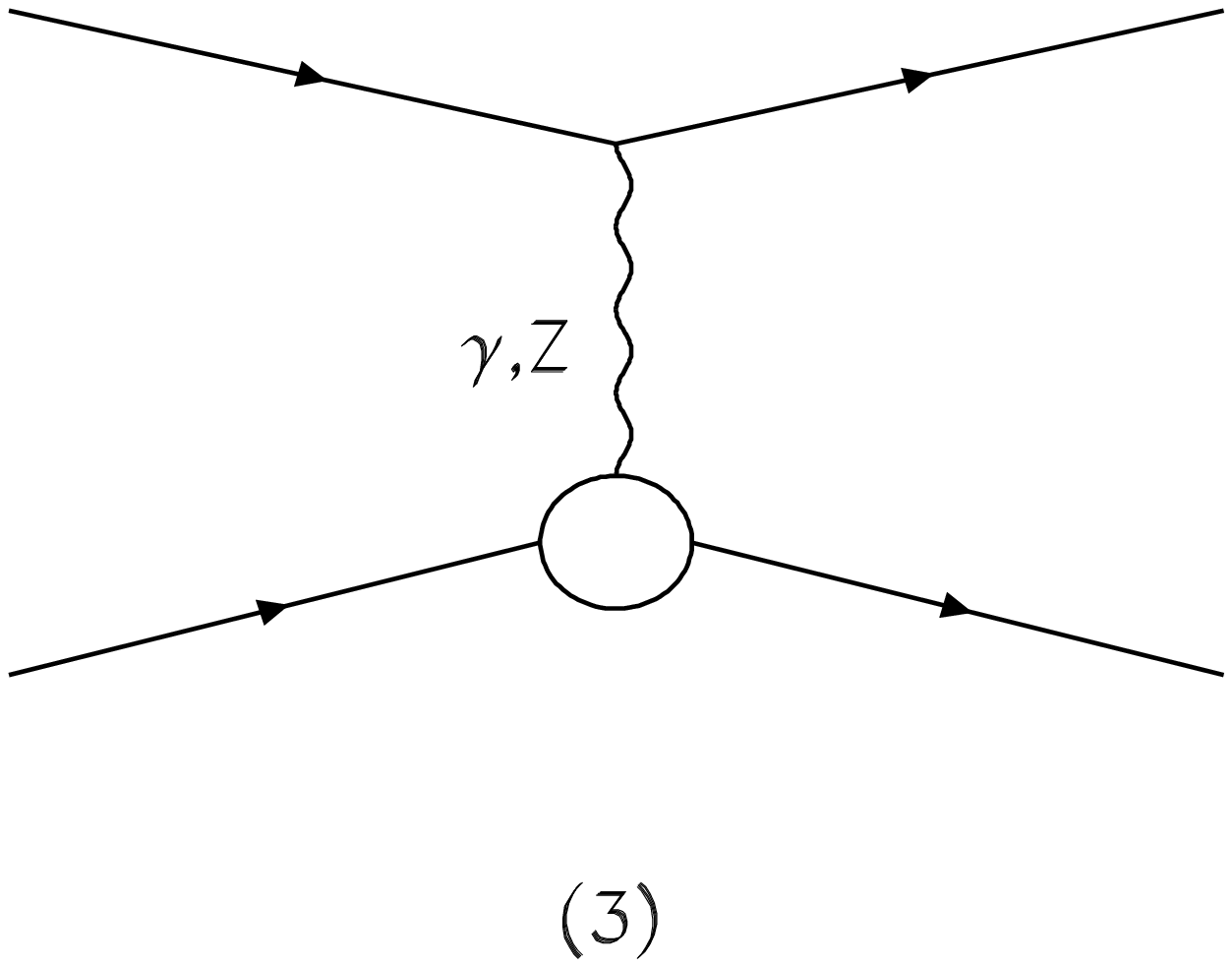} }
\end{picture}
\\[3mm]
\begin{picture}(60,60)
\put(-60,0){
\epsfxsize=5cm
\epsfysize=5cm
\epsfbox{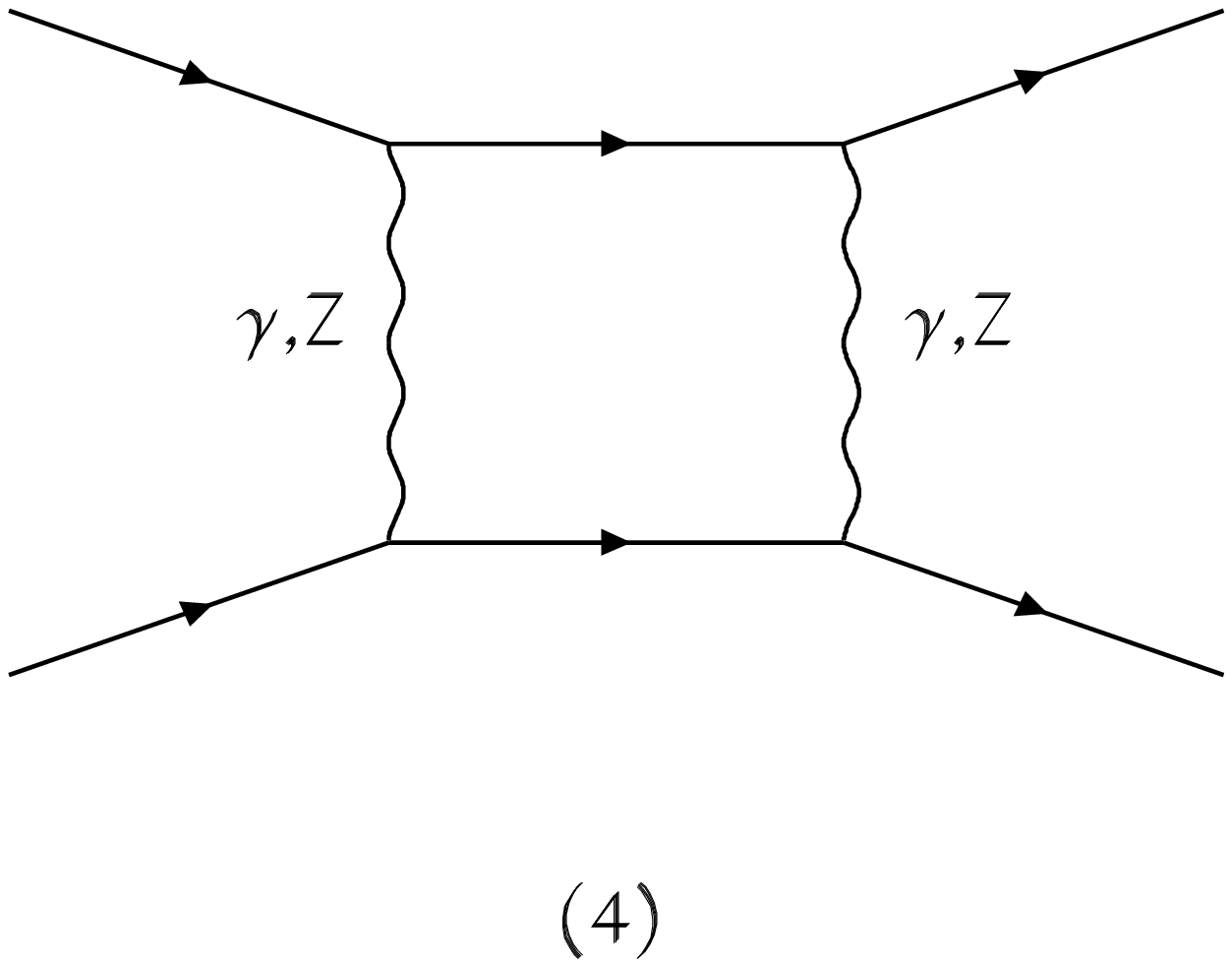} }
\end{picture}
&
\begin{picture}(60,100)
\put(-10,0){
\epsfxsize=5cm
\epsfysize=5cm
\epsfbox{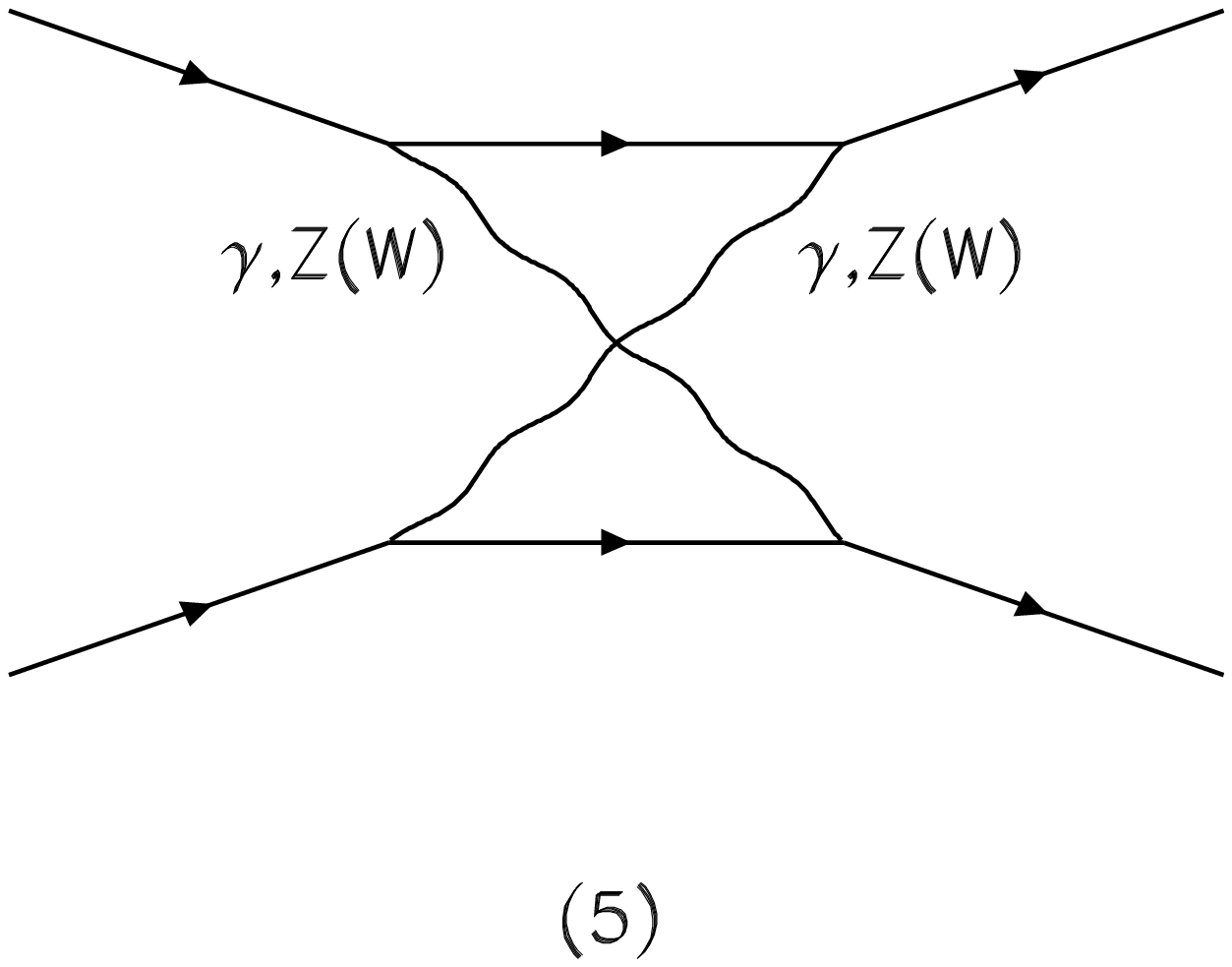} }
\end{picture}
\end{tabular}
\vspace*{-20mm}
\caption{\protect\it 
One-loop t-channel diagrams for the M{\o}ller process.
The circles represent the contributions of self-energies
and vertex functions. 
The u-channel diagrams are obtained via the interchange $k_2 \leftrightarrow p_2$. }
\label{2f}
\vspace{5mm}
\end{figure}

The contribution of virtual particles ($V$-contribution) to the observables
of Moller scattering is described by three classes
of diagrams (see Fig.~\ref{2f}): boson self-energies (BSE), vertex
functions, and two-boson exchange (boxes). In the
on-shell renormalization schemes which we
use here there are no contributions from the electron
self-energies.

The corresponding cross section is
given by
\begin{equation}
\sigma^{V}=
\sigma^{\rm BSE}+\sigma^{\rm {Ver}} +\sigma^{\rm Box},
\label{G}
\end{equation}
where the first term corresponds to the contributions
of the boson self-energies, the second term
represents the vertex diagrams, and the third term
stands for the diagrams of two-boson exchange.

The contributions of the photon and $Z$-boson
self-energies are shown symbolically in Fig.~\ref{2f}(1). 
They
do not involve infrared divergences and have the following form
\begin{equation}
\sigma^{\rm BSE}=\frac{\pi \alpha^2}{s}
 \sum_{l= 1}^8  M^S_l + (t \leftrightarrow u),
\label{S}
\end{equation}
where $M^S_l$ 
are expressed in terms of the propagators
and functions
 $M_{\rm {ev,od}}$ 
\begin{eqnarray}
M^{ijkl}_{\rm {ev}} &=&
        2 (s^2+u^2){\lambda_1}_B^{ij}{\lambda_1}_T^{kl}
       +2 (s^2-u^2){\lambda_2}_B^{ij}{\lambda_2}_T^{kl},
\nonumber \\[0.2cm]\displaystyle
M^{ijkl}_{\rm {od}} &=&
       - 2 s^2 ( {\lambda_1}_B^{ij}{\lambda_1}_T^{kl}
                +{\lambda_2}_B^{ij}{\lambda_2}_T^{kl} ),
\end{eqnarray}                                          
are defined as
\begin{eqnarray}
\displaystyle
 M^S_1 &=&   D^{\gamma t}D_S^{\gamma Z t}
        (M_{\rm ev}^{\gamma \gamma Z\gamma }+M_{\rm ev}^{Z\gamma \gamma \gamma }),\
\ \ \ \ \ \ \ \ \ \ \ \ \  M^S_2 =   -D^{\gamma u}D_S^{\gamma Z t}
        (M_{\rm od}^{\gamma \gamma Z\gamma}+M_{\rm od}^{Z\gamma \gamma \gamma}),\
\nonumber \\[0.2cm]\displaystyle
 M^S_3 &=& D^{Z t}D_S^{\gamma Z t}
        (M_{\rm ev}^{\gamma ZZZ}+M_{\rm ev}^{ZZ\gamma Z}),\
\ \ \ \ \ \ \ \ \ \ \  M^S_4 = -D^{Z u}D_S^{\gamma Z t}
        (M_{\rm od}^{\gamma ZZZ}+M_{\rm od}^{ZZ\gamma Z}),\
\nonumber \\[0.2cm]\displaystyle
 M^S_5 &=& D^{\gamma t}
        (D_S^{\gamma \gamma t}M_{\rm ev}^{\gamma \gamma \gamma \gamma} +
        D_S^{ZZ t}M_{\rm ev}^{Z\gamma Z \gamma}),\
\ \ \ \ \ \  M^S_6 = -D^{\gamma u}
        (D_S^{\gamma \gamma t}M_{\rm od}^{\gamma \gamma \gamma \gamma} +
        D_S^{ZZ t}M_{\rm od}^{Z\gamma Z \gamma}),\
\nonumber \\[0.2cm]\displaystyle
 M^S_7 &=& D^{Zt}
        (D_S^{\gamma \gamma t}M_{\rm ev}^{\gamma Z \gamma Z} +
        D_S^{ZZ t}M_{\rm ev}^{ZZZZ}),\
\ \ \ \   M^S_8 = -D^{Zu}
        (D_S^{\gamma \gamma t}M_{\rm od}^{\gamma Z \gamma Z} +
        D_S^{ZZ t}M_{\rm od}^{ZZZZ}).
\end{eqnarray}
Here,
\begin{equation}
D_S^{ijl}=-D^{il} {\hat{\Sigma}}_T^{ij}(k) D^{jl},
\label{D_S}
\end{equation}
with ${\hat{\Sigma}}_T^{ij}(k)$ 
being the transverse part of the renormalized
photon, $Z$-boson and $\gamma Z$ self-energies. 
The longitudinal parts of the boson self-energy make contributions
that are proportional to the ratios $m^2/t$
and $m^2/u$ so they are very small and are not considered here.

\subsection{ Vertices}

In order to calculate the
electron vertex corrections (2nd and 3rd diagrams in Fig.~\ref{2f}), 
we use the form factors $\delta F_{V,A}^{je}$ in manner of paper \cite{BSH86}.
Replacing the coupling
constants $v^j,\ a^j$ with these form factors 
(for example,
 $v^{\gamma (Z)} \rightarrow \delta F_V^{\gamma (Z) e}$,\ 
 $a^{\gamma (Z)} \rightarrow \delta F_A^{\gamma (Z) e}$) in the corresponding terms of the
Born functions $M_{\rm {ev,od}}$, we obtain a vertex part of the
cross sections in the form
\begin{equation}
\sigma^{\rm {Ver}}=\frac{\pi \alpha^2}{s}
 \sum_{l= 1}^4  M^V_l + (t \leftrightarrow u ),
\label{V}
\end{equation}
where
\begin{eqnarray}
\displaystyle
M^V_1&=&D^{\gamma t} ( D^{\gamma t}
            ( M_{\rm {ev}}^{ F^{\gamma}\gamma\gamma\gamma}
             +M_{\rm {ev}}^{\gamma\gamma F^{\gamma}\gamma} )
         -D^{\gamma u}
            ( M_{\rm {od}}^{ F^{\gamma}\gamma \gamma \gamma}
             +M_{\rm {od}}^{ \gamma\gamma F^{\gamma}\gamma} )),
\nonumber \\[0.cm]\displaystyle
M^V_2&=&D^{\gamma t} ( D^{Z t}
            ( M_{\rm {ev}}^{F^{\gamma}Z\gamma Z}
             +M_{\rm {ev}}^{\gamma ZF^{\gamma}Z} )
         -D^{Z u}
            ( M_{\rm {od}}^{F^{\gamma}Z\gamma Z}
             +M_{\rm {od}}^{\gamma ZF^{\gamma} Z} )),
\nonumber \\[0.cm]\displaystyle
M^V_3&=&D^{Z t} ( D^{\gamma t}
            ( M_{\rm {ev}}^{F^{Z}\gamma Z\gamma}
             +M_{\rm {ev}}^{Z\gamma F^{Z}\gamma} )
         -D^{\gamma u}
            ( M_{\rm {od}}^{F^{Z}\gamma Z \gamma }
             +M_{\rm {od}}^{Z\gamma F^Z \gamma } )),
\nonumber \\[0.cm]\displaystyle
M^V_4&=&D^{Z t} ( D^{Z t}
            ( M_{\rm {ev}}^{F^{Z}ZZZ}
             +M_{\rm {ev}}^{ZZ F^{Z}Z} )
         - D^{Z u}
            ( M_{\rm {od}}^{F^{Z}ZZZ}
             +M_{\rm {od}}^{ZZF^{Z}Z} )),
\end{eqnarray}
and
\begin{equation}
    {\lambda_V^{\rm F}}^{i,j} =
\delta F_{\rm V}^{i}v^j + \delta F_{\rm A}^{i}a^j,\
    {\lambda_A^{\rm F}}^{i,j} =
\delta F_{\rm V}^{i}a^j + \delta F_{\rm A}^{i}v^j.
\end{equation}
For the virtual photon exchange ($i=\gamma$ in Eq.(\ref{structure})), vertices look like
\begin{eqnarray}
\delta F_{\rm V}^{\gamma } & = &
  \frac{\alpha }{4\pi}
  \Bigl[ \Lambda_1
   + ({(v^Z)}^2 + {(a^Z)}^2 ) \Lambda_2(m_{Z})
   + \frac{3}{4s_{W}^2} \Lambda_3(m_{W}) \Bigr],
\label{HV1}
\end{eqnarray}
\begin{eqnarray}
\delta F_{\rm A}^{\gamma } & = &
  \frac{\alpha }{4\pi}
  \Bigl[ 2v^Za^Z \Lambda_2(m_{Z})
   + \frac{3}{4s_{W}^2} \Lambda_3(m_{W}) \Bigr],
\label{HV2}
\end{eqnarray}
and for the virtual $Z$-boson exchange  ($i=Z$ in Eq.(\ref{structure})) we have the form
\begin{eqnarray}
\delta F_{\rm V}^{Z } & = &
  \frac{\alpha }{4\pi}
  \Bigl[ v^Z \Lambda_1
   + v^Z ({(v^Z)}^2 + 3{(a^Z)}^2 ) \Lambda_2(m_{Z}) +
\nonumber
\\[0.3cm] \displaystyle
  && 
   + \frac{1}{8s_{W}^3c_{W}} \Lambda_2(m_{W})  -
 \frac{3c_{W}}{4s_{W}^3} \Lambda_3(m_{W}) \Bigr],
\label{HV3}
\end{eqnarray}
\begin{eqnarray}
\delta F_{\rm A}^{Z } & = &
  \frac{\alpha }{4\pi}
  \Bigl[ a^Z \Lambda_1
   + a^Z (3{(v^Z)}^2 + {(a^Z)}^2 ) \Lambda_2(m_{Z}) +
\nonumber 
\\[0.3cm] \displaystyle
 &&
   + \frac{1}{8s_{W}^3c_{W}} \Lambda_2(m_{W})  -
   \frac{3c_{W}}{4s_{W}^3} \Lambda_3(m_{W}) \Bigr].
\label{HV4}
\end{eqnarray}
Functions $\Lambda_1$, the contribution of triangle diagrams with additional photon, 
$\Lambda_2$, the triangle diagrams with additional massive boson -- $Z$ or $W$,  
and $\Lambda_3$, the triangle diagrams with 3-boson vertex  -- $WW\gamma$ or $WWZ$.
We denote the vertices  with an adititonal photon as light vertices (LV)
and the vertices  with an adititonal massive boson as heavy vertices (HV).
The LV terms are proportional to the function $\Lambda_1$,
and the HV terms are proportional to the combinations of functions $\Lambda_2$ and $\Lambda_3$
as it is evident from Eqs. (\ref{HV1})-(\ref{HV4}).

The contributions of separate HV, $\gamma Z$-, and $ZZ$- self energies are
strongly depend on the details of  renormalization scheme 
and for proper account of EWC they should be taken in numerical analysis 
as one gauge independent set.
For that purpose we used the on-shell renormalization scheme from Ref. \cite{Denner}.
It is of crusial importance to raise a question of dependence of EWC on details of
different renormalization schemes and this will be addressed in our following paper.

Next, we represent the vertex contribution as the sum of
infrared (IR) divergent and finite parts using the identical transformation
\begin{equation}
\sigma^{\rm Ver}=
(\sigma^{\rm Ver}- \sigma^{\rm Ver}(\lambda^2 \rightarrow s ) )
+\sigma^{\rm Ver}(\lambda^2 \rightarrow s )
= \sigma_{\rm {IR}}^{\rm Ver}
+\sigma^{\rm Ver}(\lambda^2 \rightarrow s ),
\label{b12}
\end{equation}
where $\lambda$ is the photon mass which regularizes the
infrared divergence.
Then the IR-part of vertex cross section will look like 
\begin{equation}
\sigma_{\rm {IR}}^{\rm Ver} 
=\frac{\alpha^3}{2s} \Lambda^{\rm {IR}}_1
\sum_{l= 1}^4  M^0_l + (t \leftrightarrow u),
\label{ver-IR}
\end{equation}
where
\begin{equation}
\Lambda^{\rm {IR}}_1 = 
\Lambda_1 - \Lambda_1(\lambda^2 \rightarrow s) = -2 \log\frac{s}{\lambda^2} \log\frac{-t}{em^2},
\end{equation}
$e$ is the base of the natural logarithm, and
\begin{eqnarray}
\displaystyle
   M^0_1&=&
D^{\gamma t}( D^{\gamma t}M_{\rm {ev}}^{\gamma \gamma \gamma \gamma} -
              D^{\gamma u}M_{\rm {od}}^{\gamma \gamma \gamma \gamma} ),\
\nonumber \\[0.2cm]\displaystyle
   M^0_2&=&
D^{\gamma t} ( D^{Zt}      M_{\rm {ev}}^{\gamma Z\gamma Z} -
               D^{Zu}      M_{\rm {od}}^{\gamma Z\gamma Z} ),\
\nonumber \\[0.2cm]\displaystyle
   M^0_3&=&
D^{Zt} ( D^{\gamma t}      M_{\rm {ev}}^{Z\gamma Z\gamma} -
         D^{\gamma u}      M_{\rm {od}}^{Z\gamma Z\gamma}  ) ,\
\nonumber \\[0.2cm]\displaystyle
   M^0_4&=&
D^{Zt} ( D^{Zt}M_{\rm {ev}}^{ZZZZ} -
         D^{Zu}            M_{\rm {od}}^{ZZZZ} ).
\end{eqnarray}
It is worth noticing that since the Born cross section can be presented as
\begin{equation}
\sigma^0=\frac{\pi \alpha^2}{2s}
\sum_{l= 1}^4  M^0_l + (t \leftrightarrow u)
\label{cs0-2}
\end{equation}
it is easy to make sure that in the IR-part of vertex cross section 
the Born structure is factorized with $t$- and $u$-terms separated.

\subsection{ Boxes}

In this section all formulae are written using the low energy approximation.
Our numerical estimates show the  accuracy better than 0.2\%
in the whole low energy region $0 < \sqrt{s} < 50$~GeV.
The accuracy of the approximate approach improves with the decrease of energy.
On the contrary, the calculation of boxes using FeynArts and FormCalc \cite{Hahn}
in the region $\sqrt{s} < 1$ GeV suffers from a problem of numerical instability
due to Landau singularities.
In any case, for the 11 GeV relevant for the JLab experiments, the consistency of
calculations for boxes  in both approaches is obvious and 
discrepancy has an order of $\sim 0.1$\%.

Using the identical transformation, we divide the box cross section into two parts:
\begin{equation}
\sigma^{\rm Box}
= (\sigma^{\rm Box}- \sigma^{\rm Box}(\lambda^2 \rightarrow s ) )
+\sigma^{\rm Box}(\lambda^2 \rightarrow s )
= \sigma_{\rm {IR}}^{\rm Box}
+\sigma_{\rm {F}}^{\rm Box}.
\label{box}
\end{equation}
The IR-finite box cross section looks like:
\begin{equation}
\sigma_{\rm {F}}^{\rm Box} =-\frac{\alpha^3}{s}
\Bigl( 
 \frac{ L_u^2+\pi^2}{2} 
\sum_{l=1}^4 M_l^0
+ \sum_{(ij)=1}^4 \sum_{k=\gamma,Z} B_{(ij)}^k  \Bigr)
+ (t \leftrightarrow u),
\end{equation}
where $L_u=\log(-s/u)$, the double index $(ij)$ runs like
$ (ij) = \{1,2,3,4\} = \{\gamma \gamma, \gamma Z, ZZ, WW\},$
and  expressions $B_{(ij)}^k$ take a form
\begin{eqnarray}
&&B_{(\gamma \gamma)}^k=D^{kt}
\lambda_-^{\gamma k} \delta^1_{(\gamma \gamma)} +
(D^{kt}+D^{ku})\lambda_+^{\gamma k} \delta^2_{(\gamma \gamma)},
\nonumber \\[0.3cm] \displaystyle
&&B_{(\gamma Z)}^k=D^{kt}
\lambda_-^{Z k} \delta^1_{(\gamma Z)} +
(D^{kt}+D^{ku})\lambda_+^{Z k} \delta^2_{(\gamma Z)},
\nonumber \\[0.3cm] \displaystyle
&&B_{(ZZ)}^k=D^{kt}
\lambda_-^{B k} \delta^1_{(Z Z)} +
(D^{kt}+D^{ku})\lambda_+^{B k} \delta^2_{(Z Z)},
\nonumber \\[0.3cm] \displaystyle
&&B_{(WW)}^k=D^{kt}
\lambda_-^{C k} \delta^1_{(WW)} +
(D^{kt}+D^{ku})\lambda_+^{C k} \delta^2_{(WW)}.
\end{eqnarray}
The combinations of the coupling constants are given in (\ref{b10}).
Let us recall the coupling constants for heavy boxes: 
\begin{equation}
v^B={(v^Z)}^2+{(a^Z)}^2,\ a^B=2v^Za^Z,\ v^C=a^C=1/(4s_W^2).
\end{equation}

At $s,|t|,|u| \ll m_Z^2$ the corrections $\delta_{(ij)}^{1,2}$ have a form:
\begin{eqnarray}
        \delta^1_{(\gamma \gamma)} & = &
  L_s^2(s^2+u^2)/(2t) - L_su -(L_x^2+\pi^2)u^2/t,
\nonumber \\[0.3cm] \displaystyle
        \delta^2_{(\gamma \gamma)} &=&
  L_s^2s^2/t + L_x s - (L_x^2 + \pi^2 )(s^2+u^2)/(2t),
\nonumber \\[0.3cm] \displaystyle
    \delta^1_{(\gamma Z)} &=&
  8 u^2 ( 4 I_{\gamma Z} -  {\hat I}_{\gamma Z} ),\
        \delta^2_{(\gamma Z)} =
  8 s^2 ( I_{\gamma Z} -  4 {\hat I}_{\gamma Z} ),
\nonumber \\[0.3cm] \displaystyle
        \delta^1_{(ZZ)} &=& \frac{3u^2}{2m_Z^2},\
        \delta^2_{(ZZ)} = - \frac{3s^2}{2m_Z^2},
\nonumber \\[0.3cm] \displaystyle
        \delta^1_{(WW)} &=& \frac{2u^2}{m_W^2},\
        \delta^2_{(WW)} = \frac{s^2}{2m_W^2},
\label{del-box}
\end{eqnarray}
where
\begin{equation}
L_s=\log\frac{s}{-t},\ L_x=\log\frac{u}{t},
\end{equation}
with
\begin{eqnarray}
I_{\gamma Z} &=&
\frac{1}{2 \sqrt{-u}} \int_0^1 z dz \int_0^1 dx
\frac{1}{\sqrt{\beta}}
      \log|\frac{ xz\sqrt{-u}-\sqrt{\beta} }{ xz\sqrt{-u}+\sqrt{\beta} }|,
\\[0.3cm] \displaystyle
&&  \beta = -ux^2z^2 +4(1-z)(tz(x-1)+m_Z^2).
\nonumber \\[0.3cm] 
\hat{I}_{\gamma Z} &=& I_{\gamma Z}|_{u \rightarrow -s}.
\end{eqnarray}

Finally, the IR-parts of the box cross section are given by:
\begin{eqnarray}
\sigma_{\rm {IR}}^{\rm \gamma \gamma-box} 
&=&\frac{\alpha^3}{s} \log\frac{s}{-u}\log\frac{s}{\lambda^2}
\sum_{l= 1}^2  M^0_l + (t \leftrightarrow u),
\\[0.3cm] \displaystyle
\sigma_{\rm {IR}}^{\rm \gamma Z-box} 
&=&\frac{\alpha^3}{s} \log\frac{s}{-u}\log\frac{s}{\lambda^2}
\sum_{l= 3}^4  M^0_l + (t \leftrightarrow u),
\label{ggz-IR}
\end{eqnarray}
and, summing all of the IR-terms of $V$-contribution, we have
\begin{eqnarray}
\sigma_{\rm {IR}}^{\rm Ver} &+&
\sigma_{\rm {IR}}^{\rm \gamma \gamma-box} +
\sigma_{\rm {IR}}^{\rm \gamma Z-box} 
= \frac{\alpha^3}{2s} 
(\Lambda^{\rm {IR}}_1 + 2\log\frac{s}{-u}\log\frac{s}{\lambda^2})
\sum_{l= 1}^4  M^0_l + (t \leftrightarrow u) =
\nonumber \\[0.3cm] \displaystyle
&=& -\frac{\alpha^3}{s} \log\frac{s}{\lambda^2} \log\frac{tu}{em^2s} \sum_{l= 1}^4  M^0_l + (t \leftrightarrow u)
=   -\frac{2\alpha}{\pi} \log\frac{s}{\lambda^2} \log\frac{tu}{em^2s} \sigma^0.
\label{all-IR}
\end{eqnarray}

\section{Bremsstrahlung}

\begin{figure}
\vspace{20mm}
\begin{tabular}{ccc}
\begin{picture}(60,60)
\put(-100,0){
\epsfxsize=5cm
\epsfysize=5cm
\epsfbox{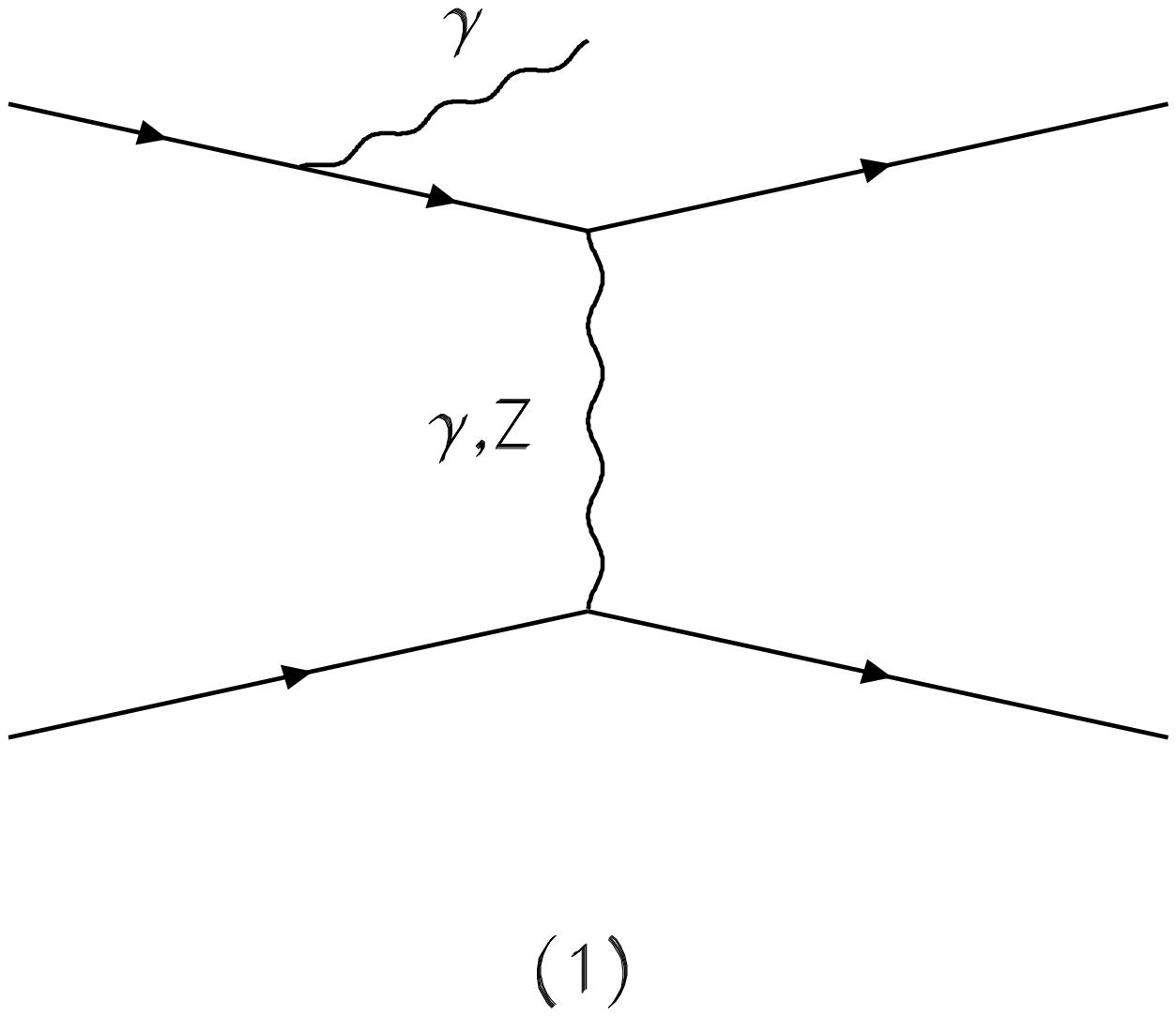} }
\end{picture}
&
\begin{picture}(60,100)
\put(-40,0){
\epsfxsize=5cm
\epsfysize=5cm
\epsfbox{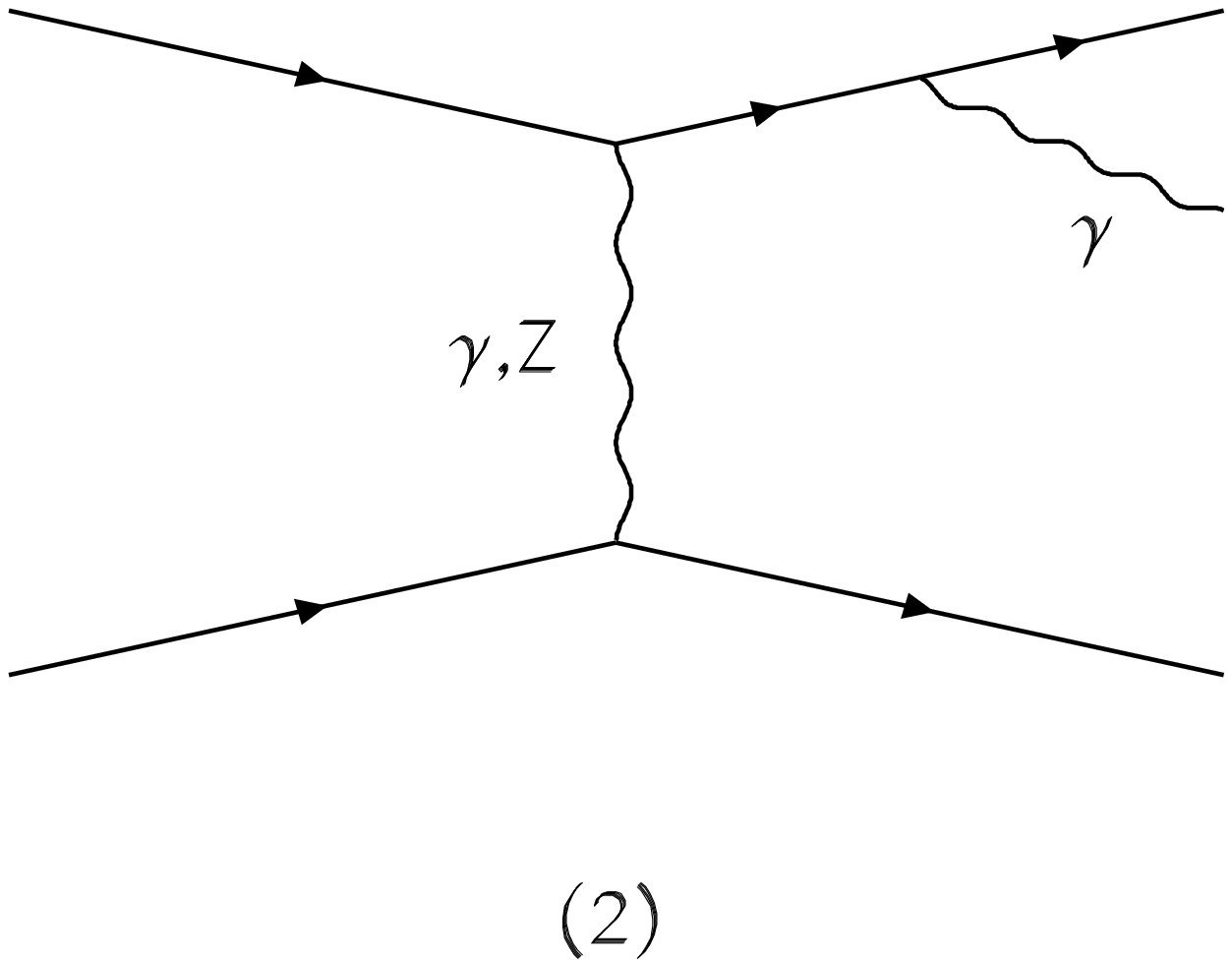} }
\end{picture}
\\[5mm]
\begin{picture}(60,100)
\put(-100,0){
\epsfxsize=5cm
\epsfysize=5cm
\epsfbox{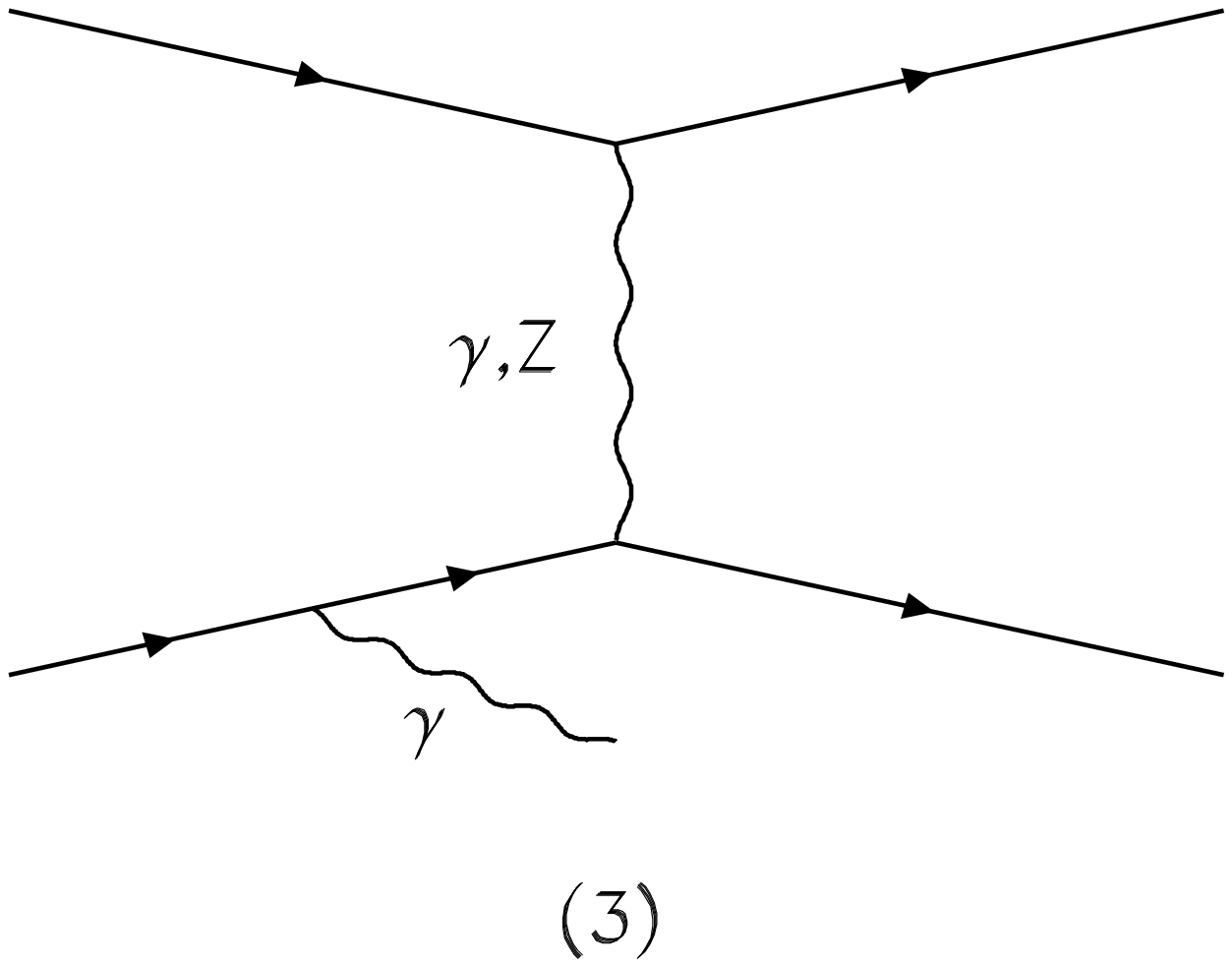} }
\end{picture}
&
\begin{picture}(60,60)
\put(-40,0){
\epsfxsize=5cm
\epsfysize=5cm
\epsfbox{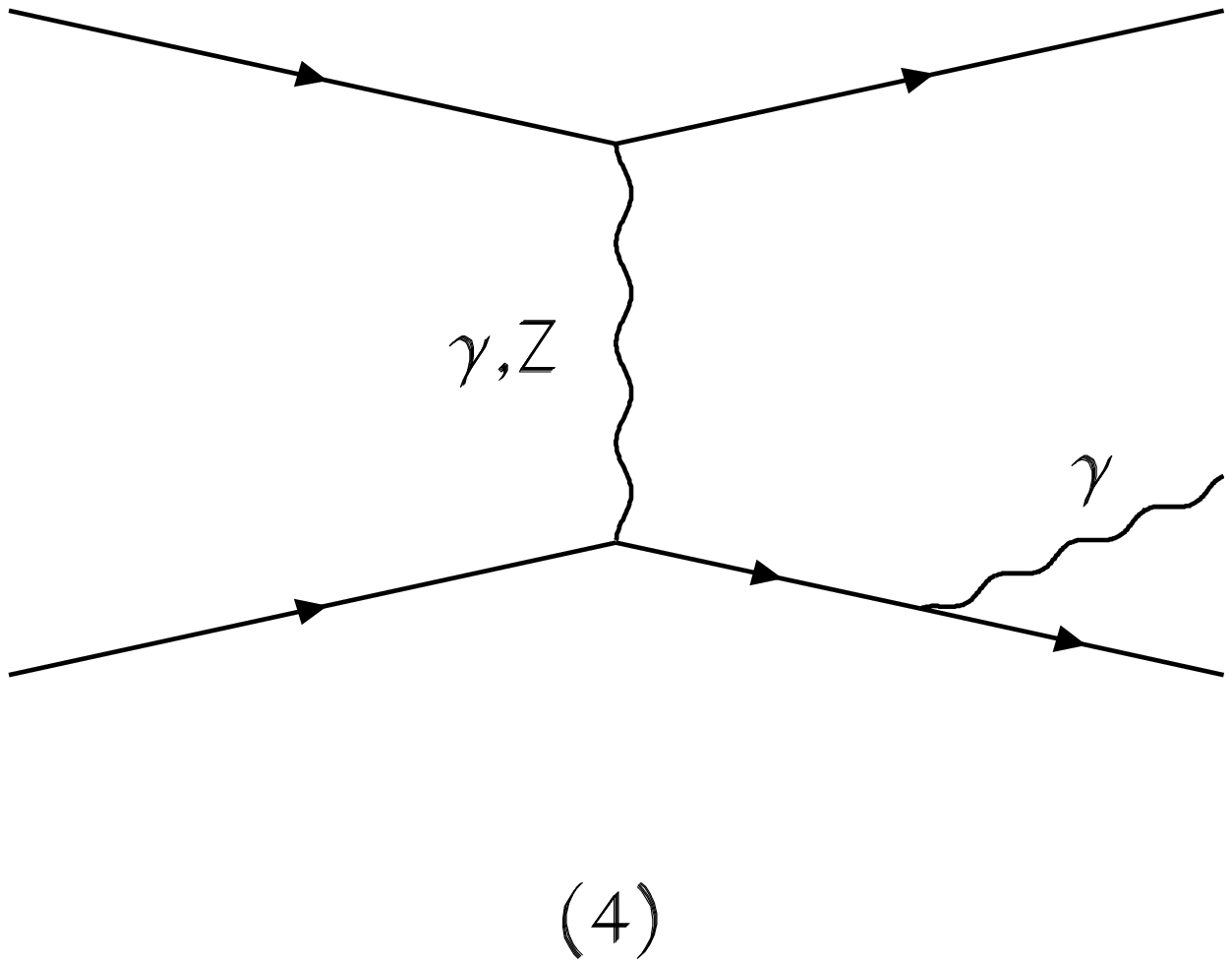} }
\end{picture}
\end{tabular}
\vspace*{-15mm}
\caption{\protect\it 
Bremsstrahlung diagrams for the M{\o}ller process in the t-channel. 
The four u-channel diagrams are obtained from those in Fig. 3 by means of
the interchange $k_2 \leftrightarrow p_2$. 
}
\label{brem}
\vspace{5mm}
\end{figure}

In order to get the IR-finite result, we have to consider the  
diagrams with the photon emission (Fig.\ref{brem}).
The whole bremsstrahlung cross section looks like 
\begin{equation}
\sigma^{R}=
\frac{\alpha^3}{4s} \int_0^{v^{\rm cut}}
\frac{s-v}{2s} dv \sum\limits_{i,j=\gamma, Z} I[M^{ij}_R],
\label{brems}
\end{equation}
where 
$v^{\rm cut} $ is the boundary of the region in Chew--Low diagram \cite{7}, and
\begin{equation}
I[M^{ij}_R]=
\frac{1}{\pi} 
 \int\frac{d^3k}{k_0} \delta((k_1+p_1-k_2-k)^2-m^2)[M^{ij}_R]
\label{I}
\end{equation}
is an integral over phase-space of an emitted photon with the  4-momentum $k$. 
The squared matrix elements corresponding to diagrams in Fig.\ref{brem} can be presented as
\begin{equation}
M^{ij}_R=(M^{it}_R-M^{iu}_R)(M^{jt}_R-M^{ju}_R)^+,
\label{MMs}
\end{equation}
where minus sign is caused by identity of final electrons and the superscripts $t (u)$ denote
the $t (u)$-channel diagrams.

Simplifying (\ref{MMs}), 
we have
\begin{equation}
M^{it}_R {M^{jt}_R}^+= \sum_{k=1,4} V^{ij}_k,\
M^{it}_R {M^{ju}_R}^+= \sum_{k=5,8} V^{ij}_k,
\label{MMs2}
\end{equation}
\begin{equation}
M^{iu}_R {M^{ju}_R}^+= M^{it}_R {M^{jt}_R}^+|_{k_2 \leftrightarrow p_2},\
M^{iu}_R {M^{jt}_R}^+= M^{it}_R {M^{ju}_R}^+|_{k_2 \leftrightarrow p_2},
\label{MMs3}
\end{equation}
where
\begin{eqnarray}
V^{ij}_1 &=&
- \mbox{Sp}
  [ G_1^{\mu \alpha} \rho^{ij}(k_1) {G_1^{\nu \alpha}}^T \Lambda(k_2) ]
  \mbox{Sp}
  [ \gamma_{\mu} \rho^{ij}(p_1) \gamma_{\nu} \Lambda(p_2) ]
  D^{it_1}D^{jt_1},
\nonumber \\
V^{ij}_2 &=&
- \mbox{Sp}
  [ G_1^{\mu \alpha} \rho^{ij}(k_1) \gamma_{\nu} \Lambda(k_2) ]
  \mbox{Sp}
  [ \gamma_{\mu} \rho^{ij}(p_1) {G_2^{\nu \alpha}}^T \Lambda(p_2) ]
  D^{it_1}D^{jt},
\nonumber \\
V^{ij}_3 &=&
- \mbox{Sp}
  [ G_2^{\mu \alpha} \rho^{ij}(p_1) \gamma_{\nu} \Lambda(p_2) ]
  \mbox{Sp}
  [ \gamma_{\mu} \rho^{ij}(k_1) {G_1^{\nu \alpha}}^T \Lambda(k_2) ]
  D^{it}D^{jt_1},
\nonumber \\
V^{ij}_4 &=&
- \mbox{Sp}
  [ G_2^{\mu \alpha} \rho^{ij}(p_1) {G_2^{\nu \alpha}}^T \Lambda(p_2) ]
  \mbox{Sp}
  [ \gamma_{\mu} \rho^{ij}(k_1) \gamma_{\nu} \Lambda(k_2) ]
  D^{it}D^{jt},
\nonumber\\ 
V^{ij}_5 &=&
- \mbox{Sp}
  [ G_1^{\mu \alpha} \rho^{ij}(k_1) {G_3^{\nu \alpha}} \Lambda(p_2)
  \gamma_{\mu} \rho^{ij}(p_1) \gamma_{\nu} \Lambda(k_2) ]
  D^{it_1}D^{ju},
\nonumber \\
V^{ij}_6 &=&
- \mbox{Sp}
  [ G_1^{\mu \alpha} \rho^{ij}(k_1) \gamma_{\nu} \Lambda(p_2)
  \gamma_{\mu} \rho^{ij}(p_1) {G_4^{\nu \alpha}} \Lambda(k_2) ]
  D^{it_1}D^{jz_2},
\nonumber\\ 
V^{ij}_7 &=&
- \mbox{Sp}
  [ \gamma_{\mu} \rho^{ij}(k_1) G_3^{\nu \alpha} \Lambda(p_2)
    G_2^{\mu \alpha} \rho^{ij}(p_1) \gamma_{\nu} \Lambda(k_2) ]
  D^{it}D^{ju},
\nonumber \\
V^{ij}_8 &=&
- \mbox{Sp}
  [ \gamma_{\mu} \rho^{ij}(k_1) \gamma_{\nu} \Lambda(p_2)
    G_2^{\mu \alpha} \rho^{ij}(p_1) G_4^{\nu \alpha} \Lambda(k_2) ]
  D^{it}D^{jz_2}.
\label{term-r}
\end{eqnarray}
We used the radiative invariants which are zero at $k \rightarrow 0 $:
\begin{equation}
 z_1=2kk_1,\ z=2kk_2,\ v_1=2kp_1,\ v=2kp_2=s+u+t-4m^2,
\end{equation}
and the invariants for propagator structure as:
\begin{equation}
t_1=(p_1-p_2)^2,\ \
z_2=(k_1-p_2)^2.
\end{equation}
Also, we applied short abbreviation for combinations with $\rho(p)$, 
density matrix of particle with 4-momentum $p$, defined as
\begin{equation}
\rho^{ij}(p)=(v^i-a^i\gamma_5)\rho(p)(v^j+a^j\gamma_5)=
\frac{1}{2}({\lambda_1}_B^{ij} \hat p -
            {\lambda_2}_B^{ij}\gamma_5 \hat p ) + {\cal O}(m).
\label{rho}
\end{equation}
We define
\begin{equation}
\Lambda(p)=\hat p + m, \ \ 
\hat p=\gamma^{\mu}p_{\mu},
\end{equation}
and
\begin{eqnarray}
 && G_1^{\mu \alpha} =
   \gamma^{\mu}\frac{2k_1^{\alpha}-\hat k \gamma^{\alpha}}{-z_1} +
   \frac{2k_2^{\alpha}+\gamma^{\alpha}\hat k}{z}\gamma^{\mu},
\nonumber \\
 && G_2^{\mu \alpha} =
   \gamma^{\mu}\frac{2p_1^{\alpha}-\hat k \gamma^{\alpha}}{-v_1} +
   \frac{2p_2^{\alpha}+\gamma^{\alpha}\hat k}{v}\gamma^{\mu},
\nonumber  \\
 && G_3^{\nu \alpha} =
 \frac{2k_1^{\alpha}-\gamma^{\alpha}\hat k}{-z_1}\gamma^{\nu} +
 \gamma^{\nu} \frac{2p_2^{\alpha}+\hat k \gamma^{\alpha}}{v},
\nonumber \\
 && G_4^{\nu \alpha} =
 \frac{2p_1^{\alpha}-\gamma^{\alpha}\hat k}{-v_1}\gamma^{\nu} +
 \gamma^{\nu} \frac{2k_2^{\alpha}+\hat k \gamma^{\alpha}}{z}.
\nonumber
\end{eqnarray}

The bremsstrahlung cross section can be broken
down into two parts (soft and hard) as 
\begin{equation}
\sigma^{R}=
\sigma_{\rm {IR}}^{R}+\sigma_{H}^{R}
\label{broke}
\end{equation}
by separating the integration domain according to $k_0 < \omega$ or $k_0 > \omega$, where
$k_0$ is the photon energy (in the reference frame co-moving
with the center of mass of primary electrons) and
$\omega$ is a parameter corresponding to the maximum soft photon energy. 
The easiest way to implement such partition
is to multiply the integrand in (\ref{brems}) by $\theta (\omega - k_0)$
and neglect photon momentum  $k \rightarrow 0$ where possible,
which would give us the soft photon cross section.

\subsection{Soft Photons and IR-divergence Cancellation}

First, we follow the methods of paper \cite{HooftVeltman} to get a well-known result (see also \cite{5-DePo,6-Pe})
for the soft photon cross section:
\begin{equation}
\sigma_{\rm {IR}}^{R}=
\frac{\alpha}{\pi} (4\log \frac{2\omega}{\lambda} \log \frac{tu}{em^2s}
-\log^2\frac{s}{em^2}+1-\frac{\pi^2}{3} +\log^2\frac{u}{t}   )
\sigma^0.
\label{IRR}
\end{equation}

Next, we summ the IR-terms of $V$- and $R$-contributions 
(formulae (\ref{all-IR}) and (\ref{IRR})),
\begin{equation}
\sigma^C=
\sigma_{\rm {IR}}^{V}+\sigma_{\rm {IR}}^R=
\frac{\alpha}{\pi} (2\log \frac{4\omega^2}{s} \log \frac{tu}{em^2s}
-\log^2\frac{s}{em^2}+1-\frac{\pi^2}{3} +\log^2\frac{u}{t}   )
\sigma^0.
\label{can}
\end{equation}
and get a result free from IR-divergence which logarithmically depends on
$\omega$ and contains $\log^2({s}/{m^2})$-terms.
Adding the contribution corresponding to $\Lambda_1$ to $\sigma^C$,  
we get an expression with the first power of collinear logarithms:
\begin{eqnarray}
\sigma^{\rm {Ver}} + \sigma^C
&\sim &\frac{ \alpha}{\pi} (\Lambda_1(\lambda^2 \rightarrow s) -\log^2\frac{s}{m^2}  ) + ...
=\frac{ \alpha}{\pi} ( \log^2\frac{-t}{m^2} - \log^2\frac{s}{m^2}  ) + ... =
\nonumber\\
&=&\frac{ \alpha}{\pi} \log\frac{-t}{s} \log\frac{-ts}{m^4} + ... .
\end{eqnarray}
with non-physical dependencies cancelled analytically.

\subsection{Hard Photons. Leading Logarithms Approach}

Now we will calculate the hard bremsstrahlung cross section
retaining in the result the leading  collinear logarithms.
This approach allows estimating the  EWC very rapidly yet provide a rather accurate result.
First let us consider exact collinear kinematics, with all relevant information 
given in Table 1 of Appendix A.
As one can see from this table, $z+v \approx (1-\eta) s$ for all peaks. 
Using the radiative invariants $k_0 = {(v + z)}/{(2\sqrt{s})}$, 
we can transform the region of integration over $v$ into 
\begin{equation}
1-2\frac{\Omega}{\sqrt{s}} < \eta < 1-2\frac{\omega}{\sqrt{s}},
\label{regi}
\end{equation}
where $\Omega$ is the maximal energy of emitted hard photon.

Now, integrating over photon phase-space and taking into consideration the results from Table 1
to simplify, we get the hard-part of bremsstrahlung cross section as
\begin{equation}
\sigma_{H}^{R}= 
\frac{\alpha^3}{4s} \int\limits_{1-2{\Omega}/{\sqrt{s}}}^{1-2{\omega}/{\sqrt{s}}}
\frac{d\eta}{1-\eta} h(\eta),
\label{HH}
\end{equation}
where
\begin{equation}
h(\eta) = h_{\rm ev,t}(\eta) + h_{\rm od,t}(\eta) + (k_2 \leftrightarrow p_2),
\label{HH2}
\end{equation}
and
\begin{eqnarray}
h_{\rm ev,t}(\eta) = \frac{s-v}{s} \Biggl( 
&& 
(2l_m-2+\frac{(1-\eta)^2}{\eta}\hat l_a) \sum_{i,j=\gamma,Z} M_{\rm ev}^{ijij}(\eta s,u)D^{it_1}D^{jt_1}|^{z_1} +
\nonumber \\[0.3cm] \displaystyle
&& 
+ (2l_m-2+ \frac{(1-\eta)^2}{\eta}l_a) \sum_{i,j=\gamma,Z} M_{\rm ev}^{ijij}(s,u/\eta)D^{it_1}D^{jt_1}|^{z} +
\nonumber \\[0.3cm] \displaystyle
&& 
+ (\hat l_a-l_s) \sum_{i,j=\gamma,Z} M_{\rm ev}^{ijij}(\eta s,u)(D^{it_1}D^{jt}+D^{it}D^{jt_1})|^{z_1} +
\nonumber \\[0.3cm] \displaystyle
&& 
+ (l_x-l_a) \sum_{i,j=\gamma,Z} M_{\rm ev}^{ijij}(s,u/\eta)(D^{it_1}D^{jt}+D^{it}D^{jt_1})|^{z} +
\nonumber \\[0.3cm] \displaystyle
&& 
+ ( (1-\eta)^2 l_u-2\eta ) \sum_{i,j=\gamma,Z} M_{\rm ev}^{ijij}(s,u) D^{it}D^{jt}|^{v_1} +
\nonumber \\[0.3cm] \displaystyle
&& 
+ 2(\frac{ (1-\eta)^2-\eta }{2-\eta} +\eta l_u) \sum_{i,j=\gamma,Z} M_{\rm ev}^{ijij}(s,u) D^{it}D^{jt}|^{v}
\Biggr),
\nonumber
\end{eqnarray}
\begin{eqnarray}
h_{\rm od,t}(\eta) = \frac{v-s}{s} \sum_{i,j=\gamma,Z} M_{\rm od}^{ijij} \Biggl( 
&& 
  (\eta(\eta^2-\eta+1)\hat l_a+\eta^2l_m-2\eta^2) D^{it_1}D^{ju}|^{z_1} 
+ (l_m-l_a) D^{it_1}D^{ju}|^{z} +
\nonumber \\[0.3cm] \displaystyle
&& 
+ \eta^2(l_m-l_s) D^{it_1}D^{jz_2}|^{z_1} 
+ (l_x+l_m+\frac{(1-\eta)^2}{\eta}l_a-2) D^{it_1}D^{jz_2}|^{z} +
\nonumber \\[0.3cm] \displaystyle
&& 
+ \eta^2(\hat l_a-l_s) D^{it}D^{ju}|^{z_1} 
+ (l_u-\frac{2}{\eta}) D^{it}D^{ju}|^{v} +
\nonumber \\[0.3cm] \displaystyle
&& 
+ ((\eta^2-\eta+1)l_u-2\eta) D^{it}D^{jz_2}|^{v_1} 
+ (l_x-l_a) D^{it}D^{jz_2}|^{z}
\Biggr).
\end{eqnarray}
The logarithms above look like 
\begin{eqnarray}
&&l_a=\log\frac{(s-v)^2}{m^2\tau},\
\hat l_a=\log\frac{(s+t)^2}{m^2\tau},\
l_m=\log\frac{-t}{m^2},\
\nonumber \\[0.3cm] \displaystyle
&& 
l_s=\log\frac{s^2}{m^4},\
l_u=\log\frac{(s+u)^2}{m^2\tau},\
l_x=\log\frac{u^2}{m^4}
\label{logs}
\end{eqnarray}
with $\tau=v+m^2$.
An operation $E|^x$ denotes the calculation of an $E$-expression in the $x$-peak.

Using the standard designation of IR-divergence extracting operation  (see a pioneer paper \cite{AP})
\begin{equation}
\int \frac{d\eta}{(1-\eta)_+} f(\eta)
=\int \frac{d\eta}{1-\eta} (f(\eta)-f(1))
\label{aapp}
\end{equation}
we can present the hard part as
\begin{equation}
\sigma_{H}^{R}= 
\sigma_{H}^{R,\Omega}+ 
\sigma_{H}^{R,\omega}= 
\frac{\alpha^3}{4s} \int\limits_{1-2\frac{\Omega}{\sqrt{s}}}^{1-2\frac{\omega}{\sqrt{s}}}
\frac{d\eta}{(1-\eta)_+} h(\eta)
+
\frac{\alpha^3}{4s} \int\limits_{1-2\frac{\Omega}{\sqrt{s}}}^{1-2\frac{\omega}{\sqrt{s}}}
\frac{d\eta}{1-\eta} h(1).
\label{HH3}
\end{equation}
Obviously, the fist term in (\ref{HH3}) is independent of $\omega$ 
(at sufficiently small $\omega$ it depends on $\Omega$ only), 
and the second term can be easily calculated as:
\begin{equation}
\sigma_{H}^{R,\omega}= 
\frac{\alpha^3}{4s} 
h(1)
\int\limits_{1-2\frac{\Omega}{\sqrt{s}}}^{1-2\frac{\omega}{\sqrt{s}}}
\frac{d\eta}{1-\eta} =
\frac{4 \alpha}{\pi} \log\frac{\Omega}{\omega} \log\frac{tu}{em^2s} \sigma^0.
\label{HH4}
\end{equation}

Finally, we obtain the desired result where the sum of all contributions
to lowest order radiative corrections is  independent of $\omega$:
\begin{equation}
\sigma_{C}+\sigma_{H}^{R,\omega}= \sigma_{C}(\omega \rightarrow \Omega).
\label{HH5}
\end{equation}

\section{Numerical analysis}

\begin{figure}
\vspace{40mm}
\hspace{10mm}
\begin{tabular}{cc}
\begin{picture}(60,60)
\put(-120,-60){
\epsfxsize=9cm
\epsfysize=9cm
\epsfbox{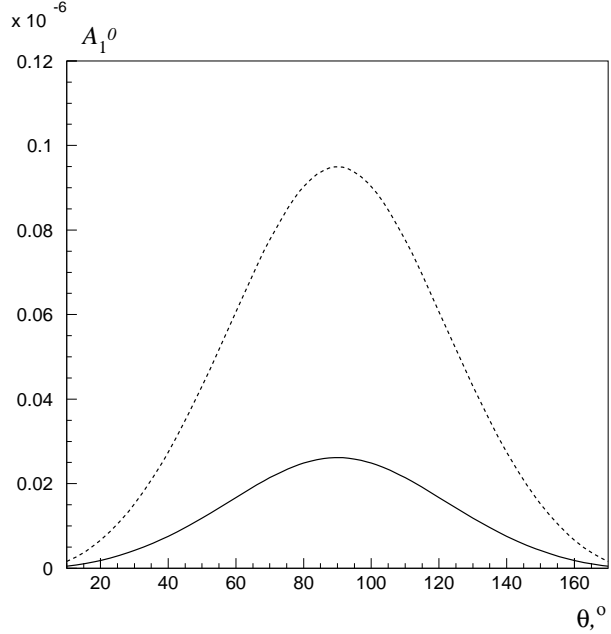} }
\end{picture}
\end{tabular}
\vspace*{15mm}
\caption{\protect\it 
Born asymmetry  versus $\theta$ at $E_{\rm lab}$ = 3.03 GeV (solid line), 
11 GeV (dashed line). }
\label{asy1}
\vspace{5mm}
\end{figure}

Fig.~\ref{asy1}  shows
the Born asymmetry $A_1^0$  versus $\theta$ at the energy of previous JLab experiment 
at $E_{\rm lab}$ = 3.03 GeV (solid line) 
and the energy of planned one with $E_{\rm lab}$ = 11 GeV (dashed line). 
Here, we used 
$\alpha=1/137.035 999$,\ 
$m_W=80.398\ \mbox{GeV}$,\  and
$m_Z=91.1876\ \mbox{GeV}$ according to \cite{PDG08}. 
It is clear that at low energies the asymmetry $A_1^0$ is proportional to $s=2mE_{\rm lab}$ 
giving $A_1^0(11 {\ \rm GeV})/A_1^0(3.03 {\ \rm GeV}) \approx  4$ for any $\theta$.

For detailed numerical calculation of EWC, we take electron, muon, and $\tau$-lepton 
masses as $m_e=0.510 998 910\ \mbox{MeV}$,  $m_\mu=0.105 658 367\ \mbox{GeV}$, $m_\tau=1.776 84\ \mbox{GeV}$ 
and quark masses for loop contributions as
$m_u=0.069 83\ \mbox{GeV}$,\ $m_c=1.2\ \mbox{GeV}$,\ $m_t=174\ \mbox{GeV}$,
$m_d=0.069 84\ \mbox{GeV}$,\ $m_s=0.15\ \mbox{GeV}$, and $m_b=4.6\ \mbox{GeV}$.
The light quark masses provide $\Delta \alpha_{had}^{(5)}(m_Z^2)$=0.02757 \cite{jeger}, 
where
\begin{equation}
\Delta \alpha_{had}^{(5)}(s)=\frac{\alpha}{\pi} \sum_{f=u,d,s,c,b} Q_f^2 \biggl(\log\frac{s}{m_f^2}-\frac{5}{3}\biggr),
\end{equation}
$Q_f$ is the electric charge of fermion $f$ in proton's charge units $q,\ (q=\sqrt{4\pi \alpha})$.
We believe that the use of the light quarks masses 
as parameters regulated by the hadron vacuum polarization
is a better choice in this case. 
Finally, for the mass of the Higg boson, we take $m_H=115\ \mbox{GeV}$. 
Although this mass is still to be determined experimentally, the dependence of EWC from $m_H$ is rather weak.


Let us define the relative corrections to the Born cross section as
$$\delta^{\rm C} = (\sigma^{\rm C}-\sigma^0)/\sigma^0,\ \rm \ C=BSE, Ver, Box, ...  $$
and to the Born asymmetry as
$$\delta^{\rm C}_A = (A_1^{\rm C}-A_1^0)/A_1^0. $$
For the convenience, we define  "{\it weak}" for relative corrections  as all BSE contributions
(including $\gamma\gamma$-SE which is not weak by nature, but we need it here 
 to account for all IR-finite contribution to asymmetry), 
HV, $ZZ$- and $WW$-boxes. This way, {\it weak} = BSE+HV+$ZZ$+$WW$.

We present all  {\it weak} and total (total = {\it weak}+QED) relative corrections 
to the unpolarized  cross section at $E_{lab}=11$ GeV and different angles $\theta$ 
with different cuts on soft ("S") 
and hard ("S+H") photon emission energy given by various $\gamma_1$ 
($\gamma_1=E_{\gamma}/\sqrt{s}$)  in Table 1 of Appendix B.
Empty cells correspond to forbidden kinematical region of this process.
If $E_{\gamma} = \omega$, we take into consideration only soft photons
and if $E_{\gamma} = \Omega$, both soft and hard photons are taken into account.
In principle, for $E_{\gamma} = \Omega$, we should treat $\omega/\sqrt{s}$ as a small parameter
but, as it was shown in Eq. (\ref{HH5}), the total result does not depend on it  in any case. 
The $\gamma\gamma$-SE contribution is small but still dominates the relative {\it weak} correction
to the unpolarized cross section. 
We can also  see a rather small difference between contributions  "S" and  "S+H".
Additionally,  we  can conclude that unpolarized cross section significantly drops with the decrease of  $\gamma_1$.

At this point, it is essential to compare the corrected parity-violating asymmetry,
which is sensitive to input parameters
and calculation scheme, with well-known existing results.
In Fig. 5 we can see the relative {\it weak} (solid line) and 
QED including soft (dashed line) corrections to the Born asymmetry $A_1^0$ 
versus $\sqrt{s}$ at $\theta$ = 90$^o$. 
In the region of high energies ($\sqrt{s} \geq $ 50 GeV) we can see an excellent
agreement with the result of Denner  and Pozzorini \cite{5-DePo} 
if we used their Standard Model parameters  (see Table below for $\delta_A^{weak}$ at different $\sqrt{s}$).

\vspace*{5mm}
\hspace*{45mm}
\begin{tabular}{|c||c|c|}
\hline
  \multicolumn{1}{|c||}{$\sqrt{s}$, GeV} 
& \multicolumn{1}{ c|}{result of Ref. \cite{5-DePo}} 
& \multicolumn{1}{ c|}{ our result} \\ 
\hline 
100   & $  -0.2787  $&$ -0.2790  $ \\
500   & $  -0.3407  $&$ -0.3406  $ \\
2000  & $  -0.9056  $&$ -0.9066  $ \\
\hline
\end{tabular}
\bigskip

Furthermore, the relative QED correction (see Fig. 8 in Ref. \cite{5-DePo} and dashed line in Fig. 5 here) 
are also in a good qualitative and numerical agreement.
In this case, we apply the same cut on the soft photon emission energy as in Ref. \cite{5-DePo} 
($\gamma_1=0.05$).

At low energy point corresponding to E-158 experiment and using our parameters set 
we find that  $\delta_A^{weak}$ = $\sim$ -54\%. 
If we translate our input parameters to the scheme by Marciano and Sirling
according to \cite{5-DePo}, we obtain  a  good agreement with the result of \cite{4-CzMa}.

All {\it  weak} and QED relative corrections to asymmetry $A_1^0$ at $\theta=90^o$ and $E_{lab}=11$ GeV
are presented in Tables 2 and 3 in Appendix B.
The $\gamma \gamma$-SE gives $\sim -0.007$ to the relative {\it weak} correction.
The $\gamma Z$-SE gives  considerable contribution and  
the $ZZ$-SE gives a contribution which is non-zero but not as large as from $\gamma Z$-SE.
It is unphysical to separate the BSE and HV due to the fact that only their sum gives a gauge invariant set. 
We used the separation  to tune up our code by comparing the values 
obtained with different renormalization schemes  \cite{BSH86} and \cite{Denner}.
We show their contribution as a total {\it weak} correction in  Table 2.
The $WW$-boxes give a rather  significant input: $\delta^{\rm WW}_A$ = 0.0238,
but on the contrary the result coming from $ZZ$-boxes is rather small and equal to 
$\delta^{\rm ZZ}_A$ = $-0.0013$.
Finally, we can see from  Table 3 that the total relative QED corrections are significant and 
strongly depend on the parameter $\gamma_1$.



\begin{figure}
\vspace{40mm}
\hspace{20mm}
\begin{tabular}{cc}
\begin{picture}(60,60)
\put(-120,-60){
\epsfxsize=9cm
\epsfysize=9cm
\epsfbox{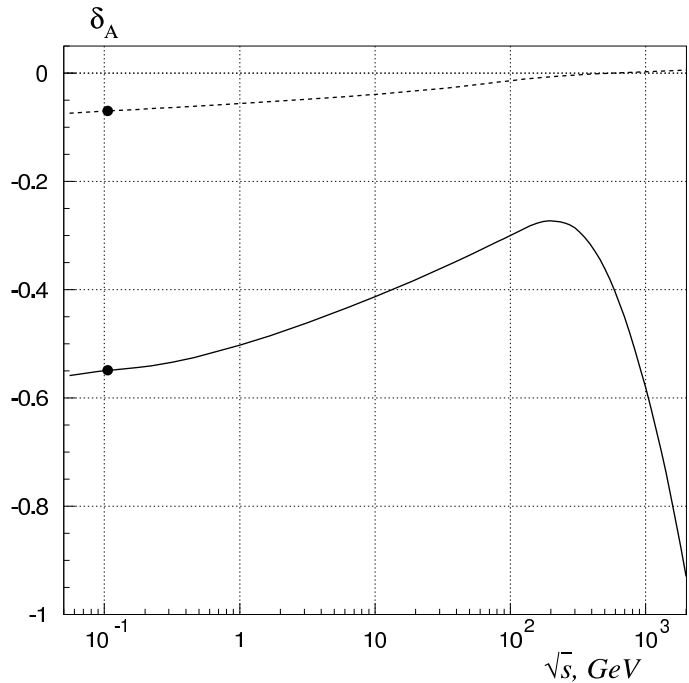} }
\end{picture}
\\\\
\end{tabular}
\vspace*{15mm}
\caption{\protect\it 
The relative weak (solid line) and  QED (dashed line) corrections 
to the Born asymmetry $A_1^0$ versus $\sqrt{s}$ at $\theta$ = 90$^o$.
The filled circle corresponds to our predictions to the future 11 GeV M{\o}ller experiment
at JLab}
\label{sverka2}
\vspace{5mm}
\end{figure}

\section{Conclusion}

M{\o}ller scattering is a very clean process that can provide indirect access to new physics at multi-TeV scales. 
The new ultra-precise measurement of the weak mixing angle via 11 GeV M{\o}ller scattering to start soon at JLab 
will require the higher order effects to be taken into account with the highest precision as well. 
In this work, we calculate the electroweak radiative corrections to asymmetry of polarized M{\o}ller 
scattering at energies relevant to the future experiments at JLab. The results are presented in both 
numerical and analytical form. As one can clearly see from our numerical data, at certain kinematic 
conditions EWC can reduce the asymmetry up to $\sim 70$\%, 
and they depend quite strongly on the experimental cuts. 
In the on-shell renormalization scheme we use, the largest contribution is coming from the 
{\it weak } corrections: self-energy graphs (especially $\gamma Z$-SE), HV, and boxes. 
The QED part is obviously very important as well. 

We believe that one of the most important result of this work, however, 
is our compact analytical and sufficiently accurate expressions. 
They can be useful for fast yet relatively precise estimations and are well suitable for building Monte Carlo generators. 
Our final results are analytically free from any non-physical parameters.  
The accuracy was controlled by comparison with the numerical 
data obtained by semi-automatic approach using FeynArts and FormCalc. 
These base languages were already successfully employed in the similar projects 
(see \cite{AA1} and \cite{AA2}, for example), so we are highly confident in their reliability. 

Since the problem of EWC for M{\o}ller scattering is rather involved, 
we believe that the tuned step-by-step comparison between different calculation approaches is essential. 
In the next work, we plan to present a detailed comparison between several calculation approaches with different 
renormalization schemes. We also plan to address the leading two-loop electroweak corrections which are very 
likely to be required by the promised experimental precision. To maximize the precision, the full set of 
one-loop EWC evaluated in this paper will also need to be re-calculated with the latest input parameters 
available at the time of the completion of the measurements, but this will be easy to do based on the 
results of the present work.

\section{ACKNOWLEDGMENTS}

We are grateful to T. Hahn, Yu. Kolomensky, E. Kuraev and J. Suarez for stimulating discussions. 
A. A. and S. B. thank the Theory Center at Jefferson Lab for hospitality in  
late 2009 when this project was inspired. 
A. I. and V. Z. thank the Acadia and Memorial Universities for hospitality in 2010. 
This work was supported by Natural Sciences and Engineering Research Council of Canada.



\ \\
\ \\

\appendix
\section {Collinear kinematics}
\label{sec:appendix-a}

Table 1. Kinematic reationships relevant to hard photon cross section.

\bigskip

\begin{center}
\begin{tabular}{|c|c|c|c|c|} 
\hline 
\multicolumn{1}{|c|}{  } & 
\multicolumn{1}{c|}{ $z_1$-peak } & 
\multicolumn{1}{c|}{ $z$-peak } & 
\multicolumn{1}{c|}{ $v_1$-peak } & 
\multicolumn{1}{c|}{ $v$-peak } \\
\hline                                                                                                                                                                
  $k$      & $(1-\eta)k_1$               &  $\frac{1-\eta}{\eta}k_2$              & $(1-\eta)p_1$                        & $\frac{1-\eta}{\eta}p_2$   \\
\hline                                                                                                                                    
  $z_1$    & $2(1-\eta)m^2\rightarrow 0$ & $\frac{1-\eta}{\eta}(2m^2-t)         $ & $(1-\eta)(s-2m^2)$                   & $(1-\eta)(s+t-4m^2+\frac{2m^2}{\eta})$    \\
\hline                                                                                                                                 
  $z$      & $(1-\eta)(2m^2-t)$          & $2\frac{1-\eta}{\eta}m^2\rightarrow 0$ & $\frac{1-\eta}{\eta}(\eta s+t-2m^2)$ & $(1-\eta)(s-2m^2)$    \\
\hline                                                                                                                                 
  $v_1$    & $(1-\eta)(s-2m^2)$          & $\frac{1-\eta}{\eta}(\eta(s-2m^2)+t) $ & $2(1-\eta)m^2\rightarrow 0$          & $(1-\eta)(2m^2-t)$    \\
\hline                                                                                                                                 
  $v$      & $(1-\eta)(s+t-2m^2)$        & $(1-\eta)(s-2m^2)$                     & $\frac{1-\eta}{\eta}(2m^2-t)$        & $2\frac{1-\eta}{\eta}m^2\rightarrow 0$    \\
\hline                                                                                                                                 
  $u$      & $ \eta(2m^2-s-t)+2m^2$      & $2m^2-t-\eta(s-2m^2)$                  & $2m^2-s+\frac{2m^2-t}{\eta}$         & $2m^2-s-t+\frac{2m^2}{\eta}$    \\
\hline                                                                                                                                 
  $k_2p_2$ & $\eta  \cdot k_1p_1$        & $\eta \cdot  k_1p_1$                   & $\eta \cdot  k_1p_1$                 & $\eta \cdot  k_1p_1$             \\
\hline                                                                                                                                                              
  $p_1p_2$ & $\eta  \cdot k_1k_2$        & $\frac{1}{\eta} \cdot k_1k_2$          & $\frac{1}{\eta} \cdot k_1k_2$        & $\eta \cdot k_1k_2$     \\
\hline                                                                                                                                                              
  $k_1p_2$ & $\frac{1}{\eta}\cdot p_1k_2$ & $\frac{1}{\eta} \cdot p_1k_2$         & $\eta \cdot p_1k_2$                  & $\eta \cdot p_1k_2$              \\
\hline                                                                                                                                    
\end{tabular}
\end{center}

\newpage

\section {Numerical analysis}
\label{sec:appendix-b}

Table 1. The unpolarized Born cross section and the relative {\it weak} and total 
corrections to it at $E_{\rm lab}$=11 GeV at different $\gamma_1$ ($\gamma_1=0.005,\ 0.01,\ 0.05$)
and $\theta$.

{\vspace*{5mm}
\begin{tabular}{|c||c||c||c|c||c|c||c|c|}
\hline
  \multicolumn{1}{|c||}{$\theta$,$^o$} 
& \multicolumn{1}{ c||}{$\sigma^0$, mb} 
& \multicolumn{1}{ c|}{\it weak} 
& \multicolumn{1}{ c|}{S, $0.005$} 
& \multicolumn{1}{ c||}{S+H, $0.005$}
& \multicolumn{1}{ c|}{S, $0.01$} 
& \multicolumn{1}{ c||}{S+H, $0.01$} 
& \multicolumn{1}{ c|}{S, $0.05$} 
& \multicolumn{1}{ c|}{S+H, $0.05$} \\
\hline 
20  & $ 0.1277\times 10^{2}  $&$   0.0087  $&$     -0.2149 $&$ -0.2148  $&$  -0.1754 $&$ -0.1758  $&$    -        $&$  -      $ \\
30  & $ 0.2607\times 10^{1}  $&$   0.0101  $&$     -0.2417 $&$ -0.2415  $&$  -0.1972 $&$ -0.1978  $&$    -        $&$  -      $ \\
40  & $ 0.8734  $&$   0.0111  $&$     -0.2595 $&$ -0.2591  $&$  -0.2118 $&$ -0.2124  $&$    -0.1012  $&$ -0.1067 $ \\
50  & $ 0.3920  $&$   0.0119  $&$     -0.2721 $&$ -0.2716  $&$  -0.2222 $&$ -0.2227  $&$    -0.1063  $&$ -0.1136 $ \\
60  & $ 0.2176  $&$   0.0126  $&$     -0.2810 $&$ -0.2805  $&$  -0.2295 $&$ -0.2303  $&$    -0.1099  $&$ -0.1183 $ \\
70  & $ 0.1444  $&$   0.0131  $&$     -0.2870 $&$ -0.2867  $&$  -0.2344 $&$ -0.2356  $&$    -0.1124  $&$ -0.1219 $ \\
80  & $ 0.1131  $&$   0.0135  $&$     -0.2905 $&$ -0.2904  $&$  -0.2373 $&$ -0.2389  $&$    -0.1139  $&$ -0.1241 $ \\
90  & $ 0.1043  $&$   0.0136  $&$     -0.2916 $&$ -0.2916  $&$  -0.2383 $&$ -0.2400  $&$    -0.1144  $&$ -0.1249 $ \\
\hline
\end{tabular}}

\ \\

Table 2. The Born asymmetry $A_1^0$ and the structure of relative {\it weak} corrections to it at $E_{\rm lab}$=11 GeV
at different $\theta$.

{\vspace*{5mm}
\begin{tabular}{|c||c|c|c|c|c|c|c|c|c|c|}
\hline
  \multicolumn{1}{|c||}{$\theta$,$^o$} 
& \multicolumn{1}{ c|}{$A_1^0$, ppb} 
& \multicolumn{1}{ c|}{$\gamma\gamma$-SE}
& \multicolumn{1}{ c|}{$\gamma Z$-SE} 
& \multicolumn{1}{ c|}{$ZZ$-SE} 
& \multicolumn{1}{ c|}{all BSE} 
& \multicolumn{1}{ c|}{$ZZ$-box} 
& \multicolumn{1}{ c|}{$WW$-box} 
& \multicolumn{1}{ c|}{HV} 
& \multicolumn{1}{ c|}{\it weak} \\
\hline
20 &  $ 6.63 $&$ -0.0043 $&$ -0.2919 $&$ -0.0105 $&$ -0.2863 $&$ -0.0013 $&$ 0.0239 $&$ -0.2946 $&$ -0.5643 $ \\  
30 &  $15.19 $&$ -0.0049 $&$ -0.2916 $&$ -0.0105 $&$ -0.3065 $&$ -0.0013 $&$ 0.0238 $&$ -0.2633 $&$ -0.5430 $ \\  
40 &  $27.45 $&$ -0.0054 $&$ -0.2914 $&$ -0.0105 $&$ -0.3051 $&$ -0.0013 $&$ 0.0238 $&$ -0.2727 $&$ -0.5508 $ \\  
50 &  $43.05 $&$ -0.0058 $&$ -0.2912 $&$ -0.0105 $&$ -0.3042 $&$ -0.0013 $&$ 0.0239 $&$ -0.2703 $&$ -0.5489 $ \\  
60 &  $60.69 $&$ -0.0062 $&$ -0.2911 $&$ -0.0105 $&$ -0.3045 $&$ -0.0013 $&$ 0.0239 $&$ -0.2714 $&$ -0.5500 $ \\  
70 &  $77.68 $&$ -0.0064 $&$ -0.2910 $&$ -0.0105 $&$ -0.3040 $&$ -0.0013 $&$ 0.0238 $&$ -0.2712 $&$ -0.5495 $ \\  
80 &  $90.28 $&$ -0.0066 $&$ -0.2909 $&$ -0.0105 $&$ -0.3040 $&$ -0.0013 $&$ 0.0238 $&$ -0.2711 $&$ -0.5493 $ \\  
90 &  $94.97 $&$ -0.0067 $&$ -0.2909 $&$ -0.0105 $&$ -0.3043 $&$ -0.0013 $&$ 0.0238 $&$ -0.2710 $&$ -0.5493 $ \\  
\hline
\end{tabular}}

\ \\
\ \\

Table 3. The Born asymmetry and the QED corrections to it at $E_{\rm lab}$=11 GeV
at different $\gamma_1$ ($\gamma_1=0.005,\ 0.01,\ 0.05$) and $\theta$.

{\vspace*{5mm}
\begin{tabular}{|c||c||c|c||c|c||c|c|}
\hline
  \multicolumn{1}{|c||}{$\theta$,$^o$} 
& \multicolumn{1}{ c||}{$A_1^0$, ppb} 
& \multicolumn{1}{ c|}{S, $0.005$}      
& \multicolumn{1}{ c||}{S+H, $0.005$}   
& \multicolumn{1}{ c|}{S, $0.01$}       
& \multicolumn{1}{ c||}{S+H, $0.01$}    
& \multicolumn{1}{ c|}{S, $0.05$}       
& \multicolumn{1}{ c|}{S+H, $0.05$} \\  
\hline 
20 & $  6.63 $&$ -0.0710 $&$  -0.0649 $&$   -0.0676 $&$  -0.0566 $&$   -        $&$  -      $ \\
30 & $ 15.19 $&$ -0.0758 $&$  -0.0736 $&$   -0.0716 $&$  -0.0686 $&$   -        $&$  -      $ \\
40 & $ 27.45 $&$ -0.0792 $&$  -0.0790 $&$   -0.0744 $&$  -0.0744 $&$   -0.0651  $&$ -0.0567 $ \\
50 & $ 43.05 $&$ -0.0817 $&$  -0.0826 $&$   -0.0763 $&$  -0.0778 $&$   -0.0663  $&$ -0.0660 $ \\
60 & $ 60.69 $&$ -0.0833 $&$  -0.0848 $&$   -0.0777 $&$  -0.0797 $&$   -0.0671  $&$ -0.0691 $ \\
70 & $ 77.68 $&$ -0.0844 $&$  -0.0859 $&$   -0.0785 $&$  -0.0805 $&$   -0.0675  $&$ -0.0701 $ \\
80 & $ 90.28 $&$ -0.0849 $&$  -0.0863 $&$   -0.0789 $&$  -0.0806 $&$   -0.0677  $&$ -0.0702 $ \\
90 & $ 94.97 $&$ -0.0850 $&$  -0.0863 $&$   -0.0790 $&$  -0.0806 $&$   -0.0678  $&$ -0.0702 $ \\
\hline
\end{tabular}}
\ \\

\end{document}